\documentclass[%
 reprint,
 superscriptaddress,
 amsmath,amssymb,
 aps,
]{revtex4-2}

\usepackage{graphicx}
\usepackage{dcolumn}
\usepackage{bm}
\usepackage{hyperref}

\usepackage{amssymb}
\usepackage{amsbsy}
\usepackage{amsmath}
\usepackage{graphicx}
\usepackage{graphics}
\usepackage{setspace}
\usepackage{array}
\usepackage{color}
\usepackage{fontenc}
\usepackage{textcomp}
\usepackage{bm}
\usepackage{float}

\newcommand{\blue}[1]{{\color{blue}{#1}}}

\begin{document}

\preprint{APS/123-QED}

\title{Altermagnetism induced surface Chern insulator}

\author{Xuance Jiang}\email{xuance@ucsb.edu
}
\altaffiliation{Present address: Department of Chemistry and Biochemistry, University of California, Santa Barbara, Santa Barbara, California 93106, USA\\ Materials Department, University of California, Santa Barbara, Santa Barbara, California 93106, USA}
\affiliation{Center for Functional Nanomaterials, Brookhaven National Laboratory, Upton, NY 11973, USA}

\affiliation{Department of Physics and Astronomy, Stony Brook University, Stony Brook, NY 11794, USA}

\author{Sayed Ali Akbar Ghorashi}\email{ghorashi@sas.upenn.edu}
\affiliation{Department of Physics and Astronomy, Stony Brook University, Stony Brook, NY 11794, USA}
\affiliation{Department of Chemistry, University of Pennsylvania, Philadelphia, USA
}

\author{Deyu Lu}
\affiliation{Center for Functional Nanomaterials, Brookhaven National Laboratory, Upton, NY 11973, USA}

\author{Jennifer Cano}\email{jennifer.cano@stonybrook.edu}
\affiliation{Department of Physics and Astronomy, Stony Brook University, Stony Brook, NY 11794, USA}
\affiliation{Center for Computational Quantum Physics, Flatiron Institute, New York, New York 10010, USA}

\date{\today}

\begin{abstract}
We propose a new pathway to the quantized anomalous Hall effect (QAHE) by coupling an altermagnet to a topological crystalline insulator (TCI). The former gaps the topological surface states of the TCI, thereby realizing the QAHE in a robust and switchable platform with near-vanishing magnetization.  We demonstrate the feasibility of this approach by studying a slab of the TCI SnTe coupled to an altermagnetic RuO$_2$	layer.  Our first-principles calculations reveal that the $d$-wave altermagnetism in RuO$_2$ induces a 7 meV gap to the Dirac surface states on the (110) surface of SnTe, producing a finite anomalous Hall effect.  
Our approach generalizes to broader classes of altermagnetic materials and TCIs, thereby providing a family of 
topological altermagnetic heterostructures with small or vanishing magnetization that support nontrivial Chern numbers. 
Our results highlight a promising new topological platform with great tunability and applications to spintronics.
\end{abstract}

\maketitle

\blue{\emph{Introduction}}.---
Materials realizing the quantum anomalous Hall effect (QAHE) are highly sought after for energy-efficient technologies exploiting their dissipationless edge states.
Conventionally, the QAHE is associated with ferromagnets; however, the scarcity of ferromagnetic insulators limits the intrinsic material platforms.
On the other hand, the first measurement of the QAHE was in a magnetically doped topological insulator \cite{chang2013experimental}, but doping introduces disorder that limits the critical temperature of the topological phase \cite{chang2023colloquium}. 
Thus, one may hope for a more robust and switchable platform in antiferromagnets,
which are less affected by stray fields and naturally exhibit  high-frequency excitations useful for spintronics \cite{jungwirth2016antiferromagnetic,baltz2018antiferromagnetic,smejkal2022anomalous}.

The emergence of altermagnets -- a new class of collinear antiferromagnets that exhibit momentum-dependent spin-splitting, despite having zero net magnetization~\cite{smejkal2022emerging,smejkal2022beyond} -- opens new avenues to explore the interplay between magnetism and topology
\cite{ghorashi2024altermagnetic,PhysRevLett.133.106701,PhysRevB.109.024404,PhysRevB.109.245306,ma2024altermagnetic,PhysRevLett.134.096703,PhysRevB.111.224406,li2025altermagnetism,ghorashi2025dynamicalgenerationhigherorderspinorbit,hadjipaschalis2025majoranas,gonzalez2025model,xie2025chiral}.
In addition, altermagnets offer great potential for electrical switching \cite{wang2024electric,duan2025antiferroelectric,chen2025electrical,gu2025ferroelectric,smejkal2024altermagnetic}.
However, previous work has focused on model Hamiltonians, while material predictions of topological altermagnets are lacking. 

Recently, we proposed a new realization of the QAHE by combining altermagnetism with topology~\cite{ghorashi2024altermagnetic}.
Specifically, an altermagnet on the surface of a three-dimensional topological insulator provides the spin-splitting needed to gap the surface state, thereby realizing the QAHE, 
with the advantage that the lack of net magnetization is favorable for realizing topological superconductivity because it does not compete with spin-singlet pairing~\cite{ghorashi2024altermagnetic,Fukaya_2025}.
However, there is a fatal subtlety that precludes a suitable material realization: all observed topological insulators \cite{hasan2010colloquium,qi2011topological,wieder2022topological} exhibit a Dirac cone at the center of the surface Brillouin zone (BZ), which is precisely where the altermagnetic splitting vanishes.

\begin{figure}
\begin{center}
\includegraphics[width=0.4\textwidth]{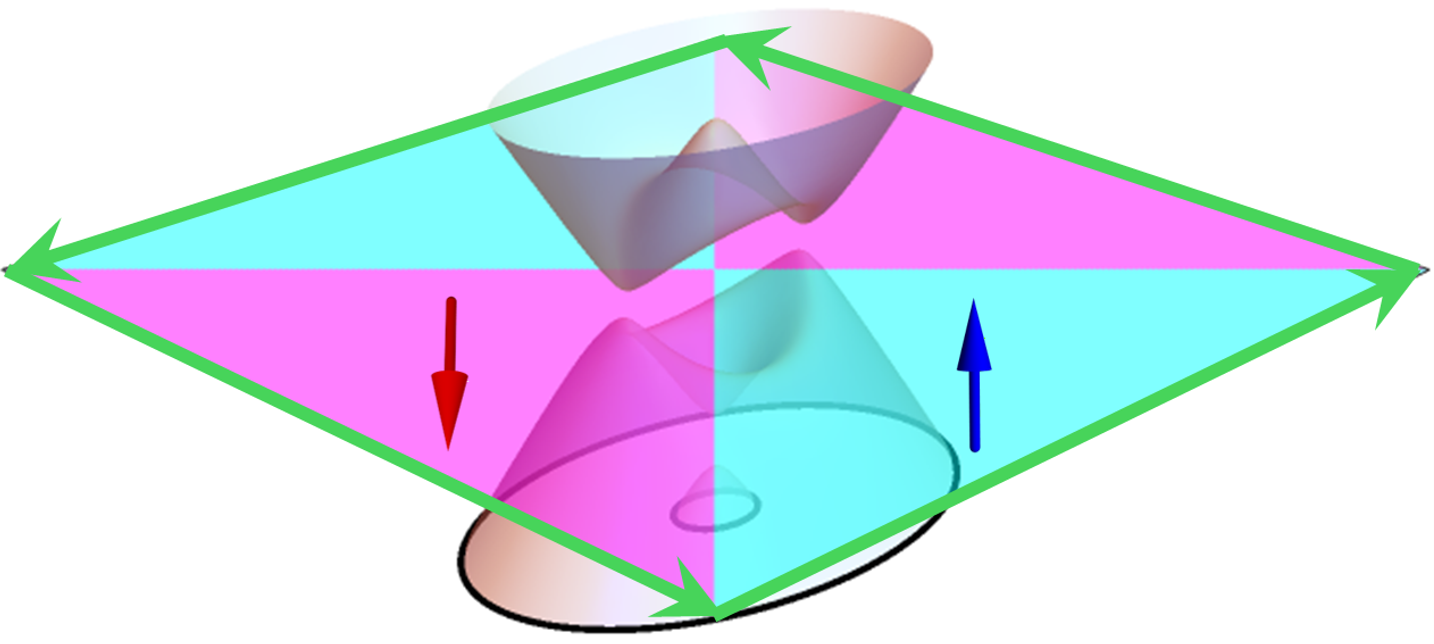}
\caption{An altermagnet on the surface of a topological crystalline insulator gaps the surface Dirac cones of the latter to realize the quantum anomalous Hall effect.
The momentum dependent spin-splitting of the altermagnet is indicated by purple/blue regions with red/blue arrows; green arrows indicate chiral edge modes.}
\label{fig:afm}
\end{center}
\end{figure}

In the present work, we propose a new heterostructure, consisting of an altermagnet on the surface of a topological \textit{crystalline} insulator (TCI).
Since the TCI is protected by crystal symmetry instead of time-reversal symmetry, its Dirac surface states do not reside at the BZ center.
Thus, proximity-induced altermagnetism can gap the topological surface states to produce the QAHE, as depicted in Fig.~\ref{fig:afm}.

As a proof-of-principle that this effect can produce measurable altermagnetic splitting, we study the interface between the prototypical altermagnet, RuO$_2$ \cite{smejkal2022emerging,smejkal2022beyond}, and the TCI SnTe \cite{hsieh2012topological,tanaka2012experimental}. 
Our results confirm the viability of inducing altermagnetism in topological surface states.
Specifically, our first-principles calculations demonstrate the formation of a 7 meV topological gap on the surface of SnTe and an accompanying spin-texture that reflects the induced altermagnetism.
To explicitly demonstrate the topological nature of the gap, we compute the layer-resolved Chern number.
Our work provides the first viable material platform realizing topological altermagnetism, opening the door to this rich field  with technological applications.




\blue{\emph{Material platform}}.--- We focus on the rutile RuO$_2$, a leading candidate room-temperature $d$-wave altermagnet: first-principles calculations predict a momentum-dependent spin splitting of order 1 eV~\cite{vsmejkal2022emerging}, and spin-resolved ARPES~\cite{lin2024observation,zhang2025observationmirroroddmirrorevenspin}, transport \cite{feng2022anomalous,PhysRevLett.130.216701,jeong2025altermagneticpolarmetallicphase} and magnetic-circular-dichroism measurements~\cite{fedchenko2024observation} have provided direct—though still debated—evidence for this unconventional altermagnetic band structure~\cite{liu2024absence,kessler2024absence,PhysRevB.109.134424,qian2025determiningnaturemagnetismaltermagnetic,PhysRevLett.132.166702}. 

Our topological ingredient is the rock-salt semiconductor SnTe, which was the first measured TCI~\cite{hsieh2012topological,tanaka2012experimental}. Mirror symmetry protects metallic surface states on its (001), (110), and (111) surfaces, while the bulk exhibits a band inversion at the four $L$ points of the fcc Brillouin zone. 

The surface termination of SnTe is important for our work.
Most studies~\cite{tanaka2012experimental,chang2016discovery,shen2014synthesis} have focused on the (001) and (111) terminations.
The latter does not align crystallographically with RuO$_2$.
We now explain why the former is also unsuitable: the (001) surface hosts four surface Dirac cones along the four equivalent $\bar{\Gamma}\bar{X}$ lines, related by rotation symmetry \cite{hsieh2012topological}.
Since the spin-splitting in the altermagnet changes sign under a four-fold rotation, when SnTe is in contact with a $d$-wave altermagnet, the magnetism could induce gaps of opposite sign on neighboring Dirac cones.
The result is a vanishing surface Chern number and, consequently, vanishing QAHE.

\begin{figure*}
\begin{center}
\includegraphics[width=0.8\textwidth]{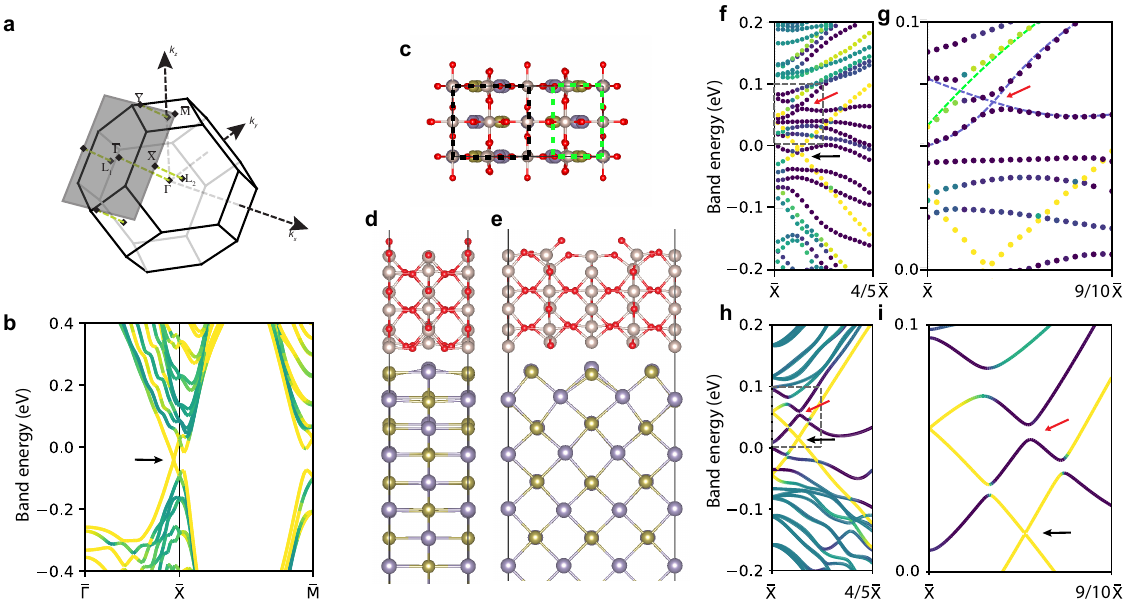}
\caption{ (a) Face-center BZ of SnTe and the surface BZ of the (110) surface. (b) Tight-binding band structure of 40 layer SnTe slab. Yellow and green color indicate the projection on the surface and bulk states, respectively, with the Dirac surface state near $\bar{X}$ indicated by the black arrow.
(c) Top view of the RuO$_2$/SnTe interface; black and green dashed rectangles indicate the RuO$_2$ and SnTe (110) unit cells. 
Side (d) and front (e) view of RuO$_2$/SnTe interface. 
(f) DFT band structure of  RuO$_2$/SnTe near $\bar{X}$ along the $\bar{X}-\bar{\Gamma}$ line. Color indicates the projection onto the top (dark purple) and bottom (yellow) layers; blue and green indicate bulk states. 
(g) Zoom-in to the gray dashed box in (f). 
Dashed lines in (g) guide the eye to avoided crossings between top surface (purple) and bulk (green) bands.
(h) The tight-binding band structure near $\bar{X}$ along the $\bar{X}-\bar{\Gamma}$ line; zoom-in in (i). 
In (f)--(i), the red arrow indicates the gapped Dirac cone on the top surface, while the black arrow indicates the gapless Dirac cone on the bottom surface.}
\label{fig:dft}
\end{center}
\end{figure*}

Thus, we now turn to the nonpolar (110) surface, which has been comparatively less explored. The (110) surface hosts two Dirac cones near the $\bar{X}$ point~\cite{hsieh2012topological}, which are related by time-reversal symmetry and protected by mirror symmetry, as shown in Fig.~\ref{fig:dft}a,b. 
In contact with an altermagnet, both Dirac cones will be gapped with the same sign, as long as the nodal line of the altermagnet is not aligned with the mirror planes.
The result is a gapped surface state with Chern number $\pm 1$ and corresponding QAHE, depicted schematically in Fig.~\ref{fig:afm}, .

This effect can be captured by a minimal model of the SnTe (110) surface. The pair of Dirac cones near $\bar{X}$ are described by~\cite{liu2013two,wang2013nontrivial},
\begin{align}
    H_{\bar{X}}=(v_x k_x s_y- v_y k_y s_x)+m\tau_x+\delta \tau_y s_y,
\end{align}
where $\tau$ and $s$ denote valley and spin space, respectively, and
$m$ and
$\delta$ describe intervalley scattering at the lattice scale. 
The crystal momenta $k_{x,y}$ are measured from $\bar{X}$ and $v_{x,y}$ are velocities in the indicated directions.
The induced $d$-wave altermagnetism is included by an additional term $J_\text{ind}\left( \cos k_x - \cos k_y \right) s_z$, which becomes $-2J_\text{ind}s_z$ upon expanding near $\bar{X}$.
This term gaps the Dirac points, leading to a quantum anomalous Hall effect with Chern number $|C| = 1$.



\blue{\emph{First-principles calculation of RuO$_2$/SnTe interface}}.---
To verify our topological argument and compute the magnitude of the induced altermagnetism,
we now provide a quantitative study of the topological gap induced at the RuO$_2$/SnTe(110) interface.

Due to the lattice mismatch of the RuO$_2$ and
SnTe (110) surfaces, we consider a slab of the heterostructure built from a $1 \times 3 \times 1$ supercell of RuO$_2$ and a $1 \times 2 \times 1$ supercell of the SnTe (110) surface as shown in Fig.~\ref{fig:dft}c-e.
The lattice mismatch in the supercell is approximately 1\% and 6\% in the $x$- and $y$- directions, respectively.

Density functional theory calculations (DFT) were performed using the projector-augmented-wave (PAW) method~\cite{kresse1999ultrasoft} implemented in the Vienna Ab initio Simulation Package (VASP)~\cite{kresse1993ab,kresse1996efficiency,kresse1996efficient}. The exchange-correlation effects are treated under the generalized gradient approximation using the Perdew-Burke-Ernzerhof (PBE) functional~\cite{perdew1996generalized} with
spin-orbit coupling included.
The electronic correlations on Ru atoms are accounted for using DFT+U~\cite{cococcioni2005linear} with $U_{\text{eff}} = 3$ eV.

We consider a slab consisting of 6 atomic layers of RuO$_2$ and 40 of SnTe, which we denote 6L-RuO$_2$/40L-SnTe.
To avoid spurious interactions between periodic images of the slab, we include a vacuum region of 20 \AA{}. The structure relaxation is performed by fixing the bottom 20 atomic layers of SnTe and relaxing the rest of the system until the total energy difference is less than 10$^{-5}$ eV and the force is less than 0.01 eV/\AA. They are unfolded to the SnTe BZ using
the Vaspbandunfolding package~\cite{zheng2025vaspbandunfolding}.

The resulting band structure of 6L-RuO$_2$/40L-SnTe is shown in Figs.~\ref{fig:dft}f,g.
The Kohn-Sham states are projected to the local atomic wave functions and colored according to their weight in the SnTe top (dark purple) and bottom (yellow) layers using a viridis color map; bulk states are green and blue.
The bottom surface states in yellow are largely unaffected by the RuO$_2$ layers on the opposite surface and the Dirac crossing remains near the Fermi level, as in a free standing SnTe slab (Fig.~\ref{fig:dft}b).

In contrast, the top surface of SnTe is strongly affected by the neighboring RuO$_2$ layers, as expected.
There are two main effects: 
first, the top surface states are shifted up approximately 0.06 eV due to a charge transfer induced interface dipole field. 
Consequently, there are some bulk states (bright green) at the same energy as the top surface states (dark purple).
The second -- and most important -- effect is that the induced altermagnetism introduces a 7 meV gap to the Dirac cone on the top surface.
The magnitude of this gap is comparable to the gap induced by ferromagnetism: for example, the SnSe/EuS interface reveals Dirac gaps of 21 meV at $\bar{\Gamma}$ and 9 meV at $\bar{M}$~\cite{yang2020topological}.

\blue{\emph{Tight-binding model for RuO$_2$/SnTe}}.---
\begin{figure}
\begin{center}
\includegraphics[width=0.5\textwidth]{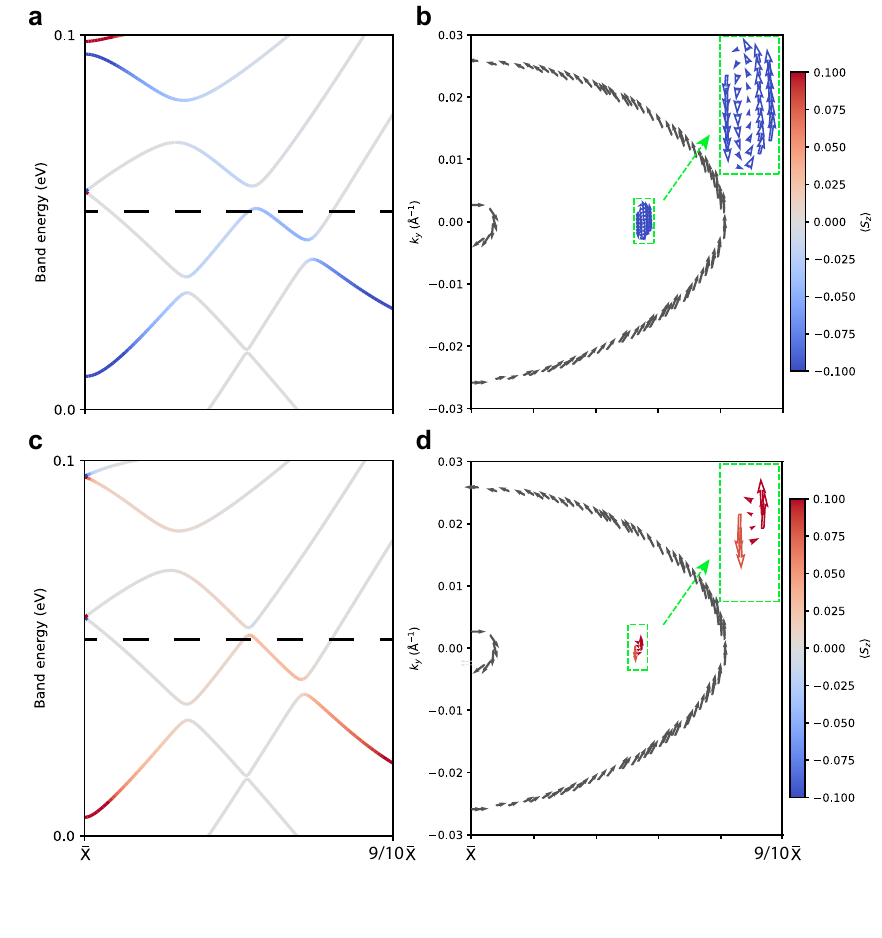 }
\caption{ (a) The tight-binding band structure near $\bar{X}$ along the $\bar{X}-\bar{\Gamma}$ line. Projection of the z-component of spin; red (blue) indicates spin up (down). (b) The spin texture at energy level $E$ indicated in (a). The arrows indicate the in-plane spin polarization and the color indicates the out-of-plane spin polarization consistent with (a). (c), (d) Same as (a),(b) after a $90^\circ$ rotation of the RuO$_2$ slab. The flipped spin polarization in (c),(d) compared to (a),(b) indicates that the slab heterostructure inherits the $d$-wave altermagnetic order in bulk RuO$_2$, even though the magnetization of the layer closest to SnTe does not change.} 
\label{fig:tbmodel}
\end{center}
\end{figure}
To directly compute the spin polarization and topological properties of the heterostructure, we construct a tight-binding (TB) model.
We first construct a TB model of a 30 unit cell thick slab of SnTe using the Wannier functions of the $p$-orbitals on the Sn and Te atoms as in Ref.~\cite{hsieh2012topological}, which is fitted from DFT calculations of bulk SnTe. To describe RuO$_2$, we use the single-orbital (Ru $d_{xy}$) two-sublattice TB model in Ref.~\cite{roig2024minimal}, applied to a three unit cell thick slab.
We couple the two slabs with hopping between the $p_{x}$ orbitals from the Te atoms at the top surface of SnTe
and the $d_{xy}$ orbitals on the bottom two layers of RuO$_2$.
An exponentially decaying potential is also added to capture the charge transfer induced interface dipole. The TB parameters that couple the two slabs, i.e., the hopping terms and the dipole field strength, are fit to the DFT band structure of the 6L-RuO$_2$/40L-SnTe heterostructure to reproduce the induced gap and shift of the DSSs shown in Fig.~\ref{fig:dft}b. The model and hopping parameters are written explicitly in the Supporting Information~\cite{supplemental2025}.


The spectrum of the TB model is shown in Figs.~\ref{fig:dft}h,i. 
As in the DFT calculation (Figs.~\ref{fig:dft}f,g), the states dominated by the top and bottom layers of SnTe are colored purple and yellow, respectively. States dominated by bulk atoms remain blue/green. 
The TB calculation reproduces the important features of the DFT calculation, namely, the dipole field shifts the surface states on the top surface of SnTe up around .05eV and the induced altermagnetism on that surface opens a 7meV gap, while the surface states on the bottom surface remain gapless.

Fig.~\ref{fig:tbmodel}a shows the induced magnetization in SnTe.
States on the bottom layer have no $z$-polarization due to the mirror symmetry that protects the Dirac cones. But on the top layer, coupling to the altermagnet breaks the mirror symmetry, which opens the gap and allows for spin polarization in the $z$-direction. 
The spin texture around the gapped Dirac cone and neighboring Fermi surfaces is shown in Fig.~\ref{fig:tbmodel}b. 

Figs.~\ref{fig:tbmodel}c,d show that when RuO$_2$ is rotated by $90^\circ$, the surface spin text flips.
This is a unique property of the altermagnetic interface. 
It demonstrates that the interface with SnTe inherits the RuO$_2$ bulk $d$-wave altermagnetic order, which flips sign upon a $90^\circ$ rotation, even though RuO$_2$ consists of stacked ferromagnetic layers and the spin polarization of the magnetic layer of RuO$_2$ at the interface does not change in the two sets of plots. 
This property is unique to coupling to an altermagnet -- it would not be present if SnTe was coupled to a ferromagnet or A-type antiferromagnet.
Note that since the four-fold rotational symmetry of RuO$_2$ is broken in the heterostructure, 
the spectra in Fig.~\ref{fig:tbmodel}a,b change slightly when compared to Fig.~\ref{fig:tbmodel}c,d.


\begin{figure}
\begin{center}
\includegraphics[width=0.5\textwidth]{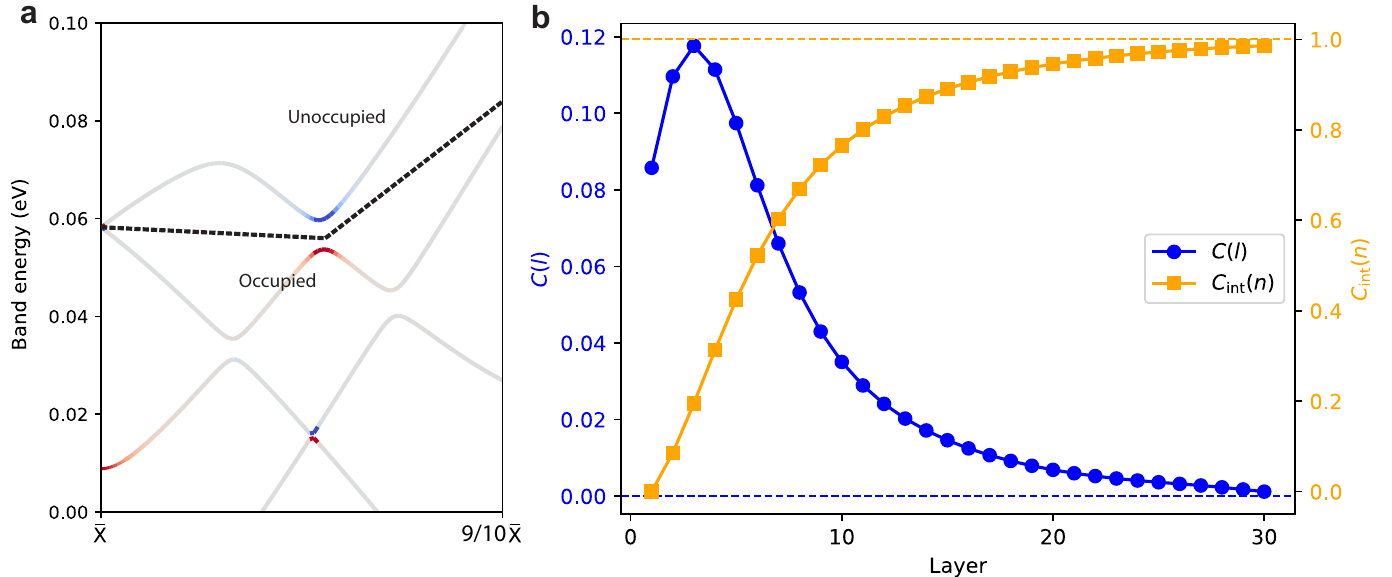}
\caption{(a) Berry curvature of the TB bands near the Dirac gap; red (blue) indicates positive (negative). 
The dashed lines indicate a continuous gap between bands (the gap at $\bar{X}$ is small but finite).
The bands below the dashed line are included in the layer-resolved Chern number calculation shown in (b); $C(l)$ (blue) indicates the partial Chern number as a function of layer $l$ from the interface, up to the middle layer, and $C_\text{int}(n) = \sum_{l<n}C(l)$ (orange) indicates its integral.}
\label{fig:lcn}
\end{center}
\end{figure}

\blue{\emph{Layered Chern number}}.---
To explicitly demonstrate the nontrivial topology at the RuO$_2$/SnTe interface, we compute the Berry curvature of the TB bands and the layer-resolved Chern number.

Fig.~\ref{fig:lcn}a shows that the gapped Dirac cone on the top surface has positive(negative) Berry curvature in the conduction(valence) band.
Since a single gapped Dirac cone contributes $e^2/2h$ to the anomalous Hall conductivity (AHC), e.g., it yields a half-quantized Hall conductivity, 
the two Dirac cones on the top surface of SnTe together contribute to an integer-quantized AHC.
However, while our DFT calculation (Fig.~\ref{fig:dft}f,g) indicates that the surface is gapped, there are bulk states that cross the interface which may cause the AHE to differ from its quantized value.

Nevertheless, a fully gapped interface may result from tuning parameters of the slab, e.g., the thickness of the layers or the strain, or from a similar heterostructure between different topological and altermagnetic layers. Thus, we now demonstrate a QAHE for an idealized related model where the topological surface is fully gapped.
Specifically, we imagine the band structure is deformed such that the states below(above) the dashed line in Fig.~\ref{fig:lcn}a are occupied(unoccupied), as indicated.
In this case, we find a quantized AHC dominated by the layers nearest the interface, as expected.

The layer-resolved AHC is computed following Ref.~\cite{varnava2018surfaces}.
The contribution of each atomic layer to the AHC is computed from the real-space Chern-density operator.
The Chern-density operator projected into the basis of tight-binding orbitals in layer $l$ gives the layer-resolved Chern number $C(l)$, which is plotted in Fig.~\ref{fig:lcn}b. 
Since the induced altermagnetism is strongest at the interface, $C(l)$ is largest near the interface, and then decays.
Integrating $C(l)$ over the slab reveals a quantized AHC, as shown in Fig.~\ref{fig:lcn}b.
The integrated value also shows that the AHC is dominated by the layers near the top surface, though is not sharply confined;
for example, in the first ten layers nearest the interface, it reaches $\sim 80\%$ of its quantized value.

\blue{\emph{Conclusion and outlook}}.---
We propose a new platform to realize the QAHE in altermagnet/TCI heterostructures. 
Our study of the RuO$_2$/SnTe interface provides the first material prediction of altermagnetic topology.
Specifically, our first-principles calculations reveal that the $d$-wave altermagnetism in RuO$_2$ opens a 7 meV gap on the two Dirac cones on the SnTe (110) surface.
The resulting gapped Dirac cones contribute to a surface anomalous Hall effect.
While charge transfer at the interface prevents the surface from being completely gapped, our work demonstrates the feasibility of this approach, bridging the gap between model Hamiltonians and the first realization of an altermagnetic QAHE.


However, realizing this heterostructure presents a challenge due to the 6\% lattice mismatch. Our work motivates a thorough investigation of similar heterostructures, e.g., between other altermagnetic materials such as MnO$_2$ and MnF$_2$ with topological mirror-Chern insulators such as SnTe or SnSe.
The search can also be expanded to broader classes of altermagnets with different order-parameter symmetries, as well as to other topological crystalline insulators. 
In particular, the materials choice should be optimized for lattice-matching and to minimize charge-transfer at the interface, so that the topological surface can be completely gapped.
Details such as slab thickness, single- vs double-side heterostructures, and strain should be systematically studied.



\begin{acknowledgments}
SAAG and JC acknowledge support from the Air Force Office of Scientific Research under Grants No. FA9550-20-1-0260 and FA9550-24-1-0222, as well as the Alfred P. Sloan
Foundation through a Sloan Research Fellowship.
JC acknowledges support from the Flatiron Institute, a division of the Simons Foundation.
This research used the Theory and Computation resources from the Center for Functional Nanomaterials (CFN), which is a U.S. Department of Energy Office of Science User Facility, at Brookhaven National Laboratory under Contract No. DE-SC0012704. This research used resources of the National Energy Research Scientific Computing Center (NERSC), a Department of Energy User Facility using NERSC award BES-ERCAP28324 and 32137.
\end{acknowledgments}

\nocite{*}

\bibliography{main}

\begin{thebibliography}{73}%
\makeatletter
\providecommand \@ifxundefined [1]{%
 \@ifx{#1\undefined}
}%
\providecommand \@ifnum [1]{%
 \ifnum #1\expandafter \@firstoftwo
 \else \expandafter \@secondoftwo
 \fi
}%
\providecommand \@ifx [1]{%
 \ifx #1\expandafter \@firstoftwo
 \else \expandafter \@secondoftwo
 \fi
}%
\providecommand \natexlab [1]{#1}%
\providecommand \enquote  [1]{``#1''}%
\providecommand \bibnamefont  [1]{#1}%
\providecommand \bibfnamefont [1]{#1}%
\providecommand \citenamefont [1]{#1}%
\providecommand \href@noop [0]{\@secondoftwo}%
\providecommand \href [0]{\begingroup \@sanitize@url \@href}%
\providecommand \@href[1]{\@@startlink{#1}\@@href}%
\providecommand \@@href[1]{\endgroup#1\@@endlink}%
\providecommand \@sanitize@url [0]{\catcode `\\12\catcode `\$12\catcode `\&12\catcode `\#12\catcode `\^12\catcode `\_12\catcode `\%12\relax}%
\providecommand \@@startlink[1]{}%
\providecommand \@@endlink[0]{}%
\providecommand \url  [0]{\begingroup\@sanitize@url \@url }%
\providecommand \@url [1]{\endgroup\@href {#1}{\urlprefix }}%
\providecommand \urlprefix  [0]{URL }%
\providecommand \Eprint [0]{\href }%
\providecommand \doibase [0]{https://doi.org/}%
\providecommand \selectlanguage [0]{\@gobble}%
\providecommand \bibinfo  [0]{\@secondoftwo}%
\providecommand \bibfield  [0]{\@secondoftwo}%
\providecommand \translation [1]{[#1]}%
\providecommand \BibitemOpen [0]{}%
\providecommand \bibitemStop [0]{}%
\providecommand \bibitemNoStop [0]{.\EOS\space}%
\providecommand \EOS [0]{\spacefactor3000\relax}%
\providecommand \BibitemShut  [1]{\csname bibitem#1\endcsname}%
\let\auto@bib@innerbib\@empty
\bibitem [{\citenamefont {Chang}\ \emph {et~al.}(2013)\citenamefont {Chang}, \citenamefont {Zhang}, \citenamefont {Feng}, \citenamefont {Shen}, \citenamefont {Zhang}, \citenamefont {Guo}, \citenamefont {Li}, \citenamefont {Ou}, \citenamefont {Wei}, \citenamefont {Wang} \emph {et~al.}}]{chang2013experimental}%
  \BibitemOpen
  \bibfield  {author} {\bibinfo {author} {\bibfnamefont {C.-Z.}\ \bibnamefont {Chang}}, \bibinfo {author} {\bibfnamefont {J.}~\bibnamefont {Zhang}}, \bibinfo {author} {\bibfnamefont {X.}~\bibnamefont {Feng}}, \bibinfo {author} {\bibfnamefont {J.}~\bibnamefont {Shen}}, \bibinfo {author} {\bibfnamefont {Z.}~\bibnamefont {Zhang}}, \bibinfo {author} {\bibfnamefont {M.}~\bibnamefont {Guo}}, \bibinfo {author} {\bibfnamefont {K.}~\bibnamefont {Li}}, \bibinfo {author} {\bibfnamefont {Y.}~\bibnamefont {Ou}}, \bibinfo {author} {\bibfnamefont {P.}~\bibnamefont {Wei}}, \bibinfo {author} {\bibfnamefont {L.-L.}\ \bibnamefont {Wang}}, \emph {et~al.},\ }\bibfield  {title} {\bibinfo {title} {Experimental observation of the quantum anomalous hall effect in a magnetic topological insulator},\ }\href@noop {} {\bibfield  {journal} {\bibinfo  {journal} {Science}\ }\textbf {\bibinfo {volume} {340}},\ \bibinfo {pages} {167} (\bibinfo {year} {2013})}\BibitemShut {NoStop}%
\bibitem [{\citenamefont {Chang}\ \emph {et~al.}(2023)\citenamefont {Chang}, \citenamefont {Liu},\ and\ \citenamefont {MacDonald}}]{chang2023colloquium}%
  \BibitemOpen
  \bibfield  {author} {\bibinfo {author} {\bibfnamefont {C.-Z.}\ \bibnamefont {Chang}}, \bibinfo {author} {\bibfnamefont {C.-X.}\ \bibnamefont {Liu}},\ and\ \bibinfo {author} {\bibfnamefont {A.~H.}\ \bibnamefont {MacDonald}},\ }\bibfield  {title} {\bibinfo {title} {Colloquium: Quantum anomalous hall effect},\ }\href {https://doi.org/10.1103/RevModPhys.95.011002} {\bibfield  {journal} {\bibinfo  {journal} {Rev. Mod. Phys.}\ }\textbf {\bibinfo {volume} {95}},\ \bibinfo {pages} {011002} (\bibinfo {year} {2023})}\BibitemShut {NoStop}%
\bibitem [{\citenamefont {Jungwirth}\ \emph {et~al.}(2016)\citenamefont {Jungwirth}, \citenamefont {Marti}, \citenamefont {Wadley},\ and\ \citenamefont {Wunderlich}}]{jungwirth2016antiferromagnetic}%
  \BibitemOpen
  \bibfield  {author} {\bibinfo {author} {\bibfnamefont {T.}~\bibnamefont {Jungwirth}}, \bibinfo {author} {\bibfnamefont {X.}~\bibnamefont {Marti}}, \bibinfo {author} {\bibfnamefont {P.}~\bibnamefont {Wadley}},\ and\ \bibinfo {author} {\bibfnamefont {J.}~\bibnamefont {Wunderlich}},\ }\bibfield  {title} {\bibinfo {title} {Antiferromagnetic spintronics},\ }\href@noop {} {\bibfield  {journal} {\bibinfo  {journal} {Nature nanotechnology}\ }\textbf {\bibinfo {volume} {11}},\ \bibinfo {pages} {231} (\bibinfo {year} {2016})}\BibitemShut {NoStop}%
\bibitem [{\citenamefont {Baltz}\ \emph {et~al.}(2018)\citenamefont {Baltz}, \citenamefont {Manchon}, \citenamefont {Tsoi}, \citenamefont {Moriyama}, \citenamefont {Ono},\ and\ \citenamefont {Tserkovnyak}}]{baltz2018antiferromagnetic}%
  \BibitemOpen
  \bibfield  {author} {\bibinfo {author} {\bibfnamefont {V.}~\bibnamefont {Baltz}}, \bibinfo {author} {\bibfnamefont {A.}~\bibnamefont {Manchon}}, \bibinfo {author} {\bibfnamefont {M.}~\bibnamefont {Tsoi}}, \bibinfo {author} {\bibfnamefont {T.}~\bibnamefont {Moriyama}}, \bibinfo {author} {\bibfnamefont {T.}~\bibnamefont {Ono}},\ and\ \bibinfo {author} {\bibfnamefont {Y.}~\bibnamefont {Tserkovnyak}},\ }\bibfield  {title} {\bibinfo {title} {Antiferromagnetic spintronics},\ }\href@noop {} {\bibfield  {journal} {\bibinfo  {journal} {Reviews of Modern Physics}\ }\textbf {\bibinfo {volume} {90}},\ \bibinfo {pages} {015005} (\bibinfo {year} {2018})}\BibitemShut {NoStop}%
\bibitem [{\citenamefont {{\v{S}}mejkal}\ \emph {et~al.}(2022{\natexlab{a}})\citenamefont {{\v{S}}mejkal}, \citenamefont {MacDonald}, \citenamefont {Sinova}, \citenamefont {Nakatsuji},\ and\ \citenamefont {Jungwirth}}]{smejkal2022anomalous}%
  \BibitemOpen
  \bibfield  {author} {\bibinfo {author} {\bibfnamefont {L.}~\bibnamefont {{\v{S}}mejkal}}, \bibinfo {author} {\bibfnamefont {A.~H.}\ \bibnamefont {MacDonald}}, \bibinfo {author} {\bibfnamefont {J.}~\bibnamefont {Sinova}}, \bibinfo {author} {\bibfnamefont {S.}~\bibnamefont {Nakatsuji}},\ and\ \bibinfo {author} {\bibfnamefont {T.}~\bibnamefont {Jungwirth}},\ }\bibfield  {title} {\bibinfo {title} {Anomalous hall antiferromagnets},\ }\href@noop {} {\bibfield  {journal} {\bibinfo  {journal} {Nature Reviews Materials}\ }\textbf {\bibinfo {volume} {7}},\ \bibinfo {pages} {482} (\bibinfo {year} {2022}{\natexlab{a}})}\BibitemShut {NoStop}%
\bibitem [{\citenamefont {{\v{S}}mejkal}\ \emph {et~al.}(2022{\natexlab{b}})\citenamefont {{\v{S}}mejkal}, \citenamefont {Sinova},\ and\ \citenamefont {Jungwirth}}]{smejkal2022emerging}%
  \BibitemOpen
  \bibfield  {author} {\bibinfo {author} {\bibfnamefont {L.}~\bibnamefont {{\v{S}}mejkal}}, \bibinfo {author} {\bibfnamefont {J.}~\bibnamefont {Sinova}},\ and\ \bibinfo {author} {\bibfnamefont {T.}~\bibnamefont {Jungwirth}},\ }\bibfield  {title} {\bibinfo {title} {Emerging research landscape of altermagnetism},\ }\href@noop {} {\bibfield  {journal} {\bibinfo  {journal} {Physical Review X}\ }\textbf {\bibinfo {volume} {12}},\ \bibinfo {pages} {040501} (\bibinfo {year} {2022}{\natexlab{b}})}\BibitemShut {NoStop}%
\bibitem [{\citenamefont {{\v{S}}mejkal}\ \emph {et~al.}(2022{\natexlab{c}})\citenamefont {{\v{S}}mejkal}, \citenamefont {Sinova},\ and\ \citenamefont {Jungwirth}}]{smejkal2022beyond}%
  \BibitemOpen
  \bibfield  {author} {\bibinfo {author} {\bibfnamefont {L.}~\bibnamefont {{\v{S}}mejkal}}, \bibinfo {author} {\bibfnamefont {J.}~\bibnamefont {Sinova}},\ and\ \bibinfo {author} {\bibfnamefont {T.}~\bibnamefont {Jungwirth}},\ }\bibfield  {title} {\bibinfo {title} {Beyond conventional ferromagnetism and antiferromagnetism: A phase with nonrelativistic spin and crystal rotation symmetry},\ }\href@noop {} {\bibfield  {journal} {\bibinfo  {journal} {Physical Review X}\ }\textbf {\bibinfo {volume} {12}},\ \bibinfo {pages} {031042} (\bibinfo {year} {2022}{\natexlab{c}})}\BibitemShut {NoStop}%
\bibitem [{\citenamefont {Ghorashi}\ \emph {et~al.}(2024)\citenamefont {Ghorashi}, \citenamefont {Hughes},\ and\ \citenamefont {Cano}}]{ghorashi2024altermagnetic}%
  \BibitemOpen
  \bibfield  {author} {\bibinfo {author} {\bibfnamefont {S.~A.~A.}\ \bibnamefont {Ghorashi}}, \bibinfo {author} {\bibfnamefont {T.~L.}\ \bibnamefont {Hughes}},\ and\ \bibinfo {author} {\bibfnamefont {J.}~\bibnamefont {Cano}},\ }\bibfield  {title} {\bibinfo {title} {Altermagnetic routes to {Majorana} modes in zero net magnetization},\ }\href {https://doi.org/10.1103/PhysRevLett.133.106601} {\bibfield  {journal} {\bibinfo  {journal} {Physical Review Letters}\ }\textbf {\bibinfo {volume} {133}},\ \bibinfo {pages} {106601} (\bibinfo {year} {2024})}\BibitemShut {NoStop}%
\bibitem [{\citenamefont {Fang}\ \emph {et~al.}(2024)\citenamefont {Fang}, \citenamefont {Cano},\ and\ \citenamefont {Ghorashi}}]{PhysRevLett.133.106701}%
  \BibitemOpen
  \bibfield  {author} {\bibinfo {author} {\bibfnamefont {Y.}~\bibnamefont {Fang}}, \bibinfo {author} {\bibfnamefont {J.}~\bibnamefont {Cano}},\ and\ \bibinfo {author} {\bibfnamefont {S.~A.~A.}\ \bibnamefont {Ghorashi}},\ }\bibfield  {title} {\bibinfo {title} {Quantum geometry induced nonlinear transport in altermagnets},\ }\href {https://doi.org/10.1103/PhysRevLett.133.106701} {\bibfield  {journal} {\bibinfo  {journal} {Phys. Rev. Lett.}\ }\textbf {\bibinfo {volume} {133}},\ \bibinfo {pages} {106701} (\bibinfo {year} {2024})}\BibitemShut {NoStop}%
\bibitem [{\citenamefont {Fernandes}\ \emph {et~al.}(2024)\citenamefont {Fernandes}, \citenamefont {de~Carvalho}, \citenamefont {Birol},\ and\ \citenamefont {Pereira}}]{PhysRevB.109.024404}%
  \BibitemOpen
  \bibfield  {author} {\bibinfo {author} {\bibfnamefont {R.~M.}\ \bibnamefont {Fernandes}}, \bibinfo {author} {\bibfnamefont {V.~S.}\ \bibnamefont {de~Carvalho}}, \bibinfo {author} {\bibfnamefont {T.}~\bibnamefont {Birol}},\ and\ \bibinfo {author} {\bibfnamefont {R.~G.}\ \bibnamefont {Pereira}},\ }\bibfield  {title} {\bibinfo {title} {Topological transition from nodal to nodeless {Zeeman} splitting in altermagnets},\ }\href {https://doi.org/10.1103/PhysRevB.109.024404} {\bibfield  {journal} {\bibinfo  {journal} {Phys. Rev. B}\ }\textbf {\bibinfo {volume} {109}},\ \bibinfo {pages} {024404} (\bibinfo {year} {2024})}\BibitemShut {NoStop}%
\bibitem [{\citenamefont {Ezawa}(2024)}]{PhysRevB.109.245306}%
  \BibitemOpen
  \bibfield  {author} {\bibinfo {author} {\bibfnamefont {M.}~\bibnamefont {Ezawa}},\ }\bibfield  {title} {\bibinfo {title} {Detecting the {N\'eel} vector of altermagnets in heterostructures with a topological insulator and a crystalline valley-edge insulator},\ }\href {https://doi.org/10.1103/PhysRevB.109.245306} {\bibfield  {journal} {\bibinfo  {journal} {Phys. Rev. B}\ }\textbf {\bibinfo {volume} {109}},\ \bibinfo {pages} {245306} (\bibinfo {year} {2024})}\BibitemShut {NoStop}%
\bibitem [{\citenamefont {Ma}\ and\ \citenamefont {Jia}(2024)}]{ma2024altermagnetic}%
  \BibitemOpen
  \bibfield  {author} {\bibinfo {author} {\bibfnamefont {H.-Y.}\ \bibnamefont {Ma}}\ and\ \bibinfo {author} {\bibfnamefont {J.-F.}\ \bibnamefont {Jia}},\ }\bibfield  {title} {\bibinfo {title} {Altermagnetic topological insulator and the selection rules},\ }\href@noop {} {\bibfield  {journal} {\bibinfo  {journal} {Physical Review B}\ }\textbf {\bibinfo {volume} {110}},\ \bibinfo {pages} {064426} (\bibinfo {year} {2024})}\BibitemShut {NoStop}%
\bibitem [{\citenamefont {Antonenko}\ \emph {et~al.}(2025)\citenamefont {Antonenko}, \citenamefont {Fernandes},\ and\ \citenamefont {Venderbos}}]{PhysRevLett.134.096703}%
  \BibitemOpen
  \bibfield  {author} {\bibinfo {author} {\bibfnamefont {D.~S.}\ \bibnamefont {Antonenko}}, \bibinfo {author} {\bibfnamefont {R.~M.}\ \bibnamefont {Fernandes}},\ and\ \bibinfo {author} {\bibfnamefont {J.~W.~F.}\ \bibnamefont {Venderbos}},\ }\bibfield  {title} {\bibinfo {title} {Mirror {Chern} bands and {Weyl} nodal loops in altermagnets},\ }\href {https://doi.org/10.1103/PhysRevLett.134.096703} {\bibfield  {journal} {\bibinfo  {journal} {Phys. Rev. Lett.}\ }\textbf {\bibinfo {volume} {134}},\ \bibinfo {pages} {096703} (\bibinfo {year} {2025})}\BibitemShut {NoStop}%
\bibitem [{\citenamefont {Parshukov}\ \emph {et~al.}(2025)\citenamefont {Parshukov}, \citenamefont {Wiedmann},\ and\ \citenamefont {Schnyder}}]{PhysRevB.111.224406}%
  \BibitemOpen
  \bibfield  {author} {\bibinfo {author} {\bibfnamefont {K.}~\bibnamefont {Parshukov}}, \bibinfo {author} {\bibfnamefont {R.}~\bibnamefont {Wiedmann}},\ and\ \bibinfo {author} {\bibfnamefont {A.~P.}\ \bibnamefont {Schnyder}},\ }\bibfield  {title} {\bibinfo {title} {Topological crossings in two-dimensional altermagnets: Symmetry classification and topological responses},\ }\href {https://doi.org/10.1103/PhysRevB.111.224406} {\bibfield  {journal} {\bibinfo  {journal} {Phys. Rev. B}\ }\textbf {\bibinfo {volume} {111}},\ \bibinfo {pages} {224406} (\bibinfo {year} {2025})}\BibitemShut {NoStop}%
\bibitem [{\citenamefont {Li}\ \emph {et~al.}(2025)\citenamefont {Li}, \citenamefont {Li},\ and\ \citenamefont {Qiao}}]{li2025altermagnetism}%
  \BibitemOpen
  \bibfield  {author} {\bibinfo {author} {\bibfnamefont {Z.}~\bibnamefont {Li}}, \bibinfo {author} {\bibfnamefont {Z.}~\bibnamefont {Li}},\ and\ \bibinfo {author} {\bibfnamefont {Z.}~\bibnamefont {Qiao}},\ }\bibfield  {title} {\bibinfo {title} {Altermagnetism-induced topological phase transitions in the {Kane-Mele} model},\ }\href@noop {} {\bibfield  {journal} {\bibinfo  {journal} {Physical Review B}\ }\textbf {\bibinfo {volume} {111}},\ \bibinfo {pages} {155303} (\bibinfo {year} {2025})}\BibitemShut {NoStop}%
\bibitem [{\citenamefont {Ghorashi}\ and\ \citenamefont {Li}(2025)}]{ghorashi2025dynamicalgenerationhigherorderspinorbit}%
  \BibitemOpen
  \bibfield  {author} {\bibinfo {author} {\bibfnamefont {S.~A.~A.}\ \bibnamefont {Ghorashi}}\ and\ \bibinfo {author} {\bibfnamefont {Q.}~\bibnamefont {Li}},\ }\href {https://arxiv.org/abs/2504.00122} {\bibinfo {title} {Dynamical generation of higher-order spin-orbit couplings, topology and persistent spin texture in light-irradiated altermagnets}} (\bibinfo {year} {2025}),\ \Eprint {https://arxiv.org/abs/2504.00122} {arXiv:2504.00122 [cond-mat.mes-hall]} \BibitemShut {NoStop}%
\bibitem [{\citenamefont {Hadjipaschalis}\ \emph {et~al.}(2025)\citenamefont {Hadjipaschalis}, \citenamefont {Ghorashi},\ and\ \citenamefont {Cano}}]{hadjipaschalis2025majoranas}%
  \BibitemOpen
  \bibfield  {author} {\bibinfo {author} {\bibfnamefont {A.}~\bibnamefont {Hadjipaschalis}}, \bibinfo {author} {\bibfnamefont {S.~A.~A.}\ \bibnamefont {Ghorashi}},\ and\ \bibinfo {author} {\bibfnamefont {J.}~\bibnamefont {Cano}},\ }\bibfield  {title} {\bibinfo {title} {Majoranas with a twist: Tunable {Majorana} zero modes in altermagnetic heterostructures},\ }\href@noop {} {\bibfield  {journal} {\bibinfo  {journal} {arXiv preprint arXiv:2507.00119}\ } (\bibinfo {year} {2025})}\BibitemShut {NoStop}%
\bibitem [{\citenamefont {Gonzalez-Hernandez}\ and\ \citenamefont {Uribe}(2025)}]{gonzalez2025model}%
  \BibitemOpen
  \bibfield  {author} {\bibinfo {author} {\bibfnamefont {R.}~\bibnamefont {Gonzalez-Hernandez}}\ and\ \bibinfo {author} {\bibfnamefont {B.}~\bibnamefont {Uribe}},\ }\bibfield  {title} {\bibinfo {title} {Model hamiltonian for altermagnetic topological insulators},\ }\href@noop {} {\bibfield  {journal} {\bibinfo  {journal} {arXiv preprint arXiv:2507.23173}\ } (\bibinfo {year} {2025})}\BibitemShut {NoStop}%
\bibitem [{\citenamefont {Xie}\ \emph {et~al.}(2025)\citenamefont {Xie}, \citenamefont {Guo}, \citenamefont {Wang}, \citenamefont {Meng}, \citenamefont {Wang}, \citenamefont {Cheng},\ and\ \citenamefont {Zhou}}]{xie2025chiral}%
  \BibitemOpen
  \bibfield  {author} {\bibinfo {author} {\bibfnamefont {C.}~\bibnamefont {Xie}}, \bibinfo {author} {\bibfnamefont {Z.}~\bibnamefont {Guo}}, \bibinfo {author} {\bibfnamefont {W.}~\bibnamefont {Wang}}, \bibinfo {author} {\bibfnamefont {W.}~\bibnamefont {Meng}}, \bibinfo {author} {\bibfnamefont {X.}~\bibnamefont {Wang}}, \bibinfo {author} {\bibfnamefont {Z.}~\bibnamefont {Cheng}},\ and\ \bibinfo {author} {\bibfnamefont {X.}~\bibnamefont {Zhou}},\ }\bibfield  {title} {\bibinfo {title} {Chiral altermagnetic second-order topological phases and sign-reversible transport},\ }\href@noop {} {\bibfield  {journal} {\bibinfo  {journal} {arXiv preprint arXiv:2508.12770}\ } (\bibinfo {year} {2025})}\BibitemShut {NoStop}%
\bibitem [{\citenamefont {Wang}\ \emph {et~al.}(2024)\citenamefont {Wang}, \citenamefont {Wang}, \citenamefont {Liu}, \citenamefont {Zhang},\ and\ \citenamefont {Zhang}}]{wang2024electric}%
  \BibitemOpen
  \bibfield  {author} {\bibinfo {author} {\bibfnamefont {D.}~\bibnamefont {Wang}}, \bibinfo {author} {\bibfnamefont {H.}~\bibnamefont {Wang}}, \bibinfo {author} {\bibfnamefont {L.}~\bibnamefont {Liu}}, \bibinfo {author} {\bibfnamefont {J.}~\bibnamefont {Zhang}},\ and\ \bibinfo {author} {\bibfnamefont {H.}~\bibnamefont {Zhang}},\ }\bibfield  {title} {\bibinfo {title} {Electric-field-induced switchable two-dimensional altermagnets},\ }\href@noop {} {\bibfield  {journal} {\bibinfo  {journal} {Nano Letters}\ }\textbf {\bibinfo {volume} {25}},\ \bibinfo {pages} {498} (\bibinfo {year} {2024})}\BibitemShut {NoStop}%
\bibitem [{\citenamefont {Duan}\ \emph {et~al.}(2025)\citenamefont {Duan}, \citenamefont {Zhang}, \citenamefont {Zhu}, \citenamefont {Liu}, \citenamefont {Zhang}, \citenamefont {{\v{Z}}uti{\'c}},\ and\ \citenamefont {Zhou}}]{duan2025antiferroelectric}%
  \BibitemOpen
  \bibfield  {author} {\bibinfo {author} {\bibfnamefont {X.}~\bibnamefont {Duan}}, \bibinfo {author} {\bibfnamefont {J.}~\bibnamefont {Zhang}}, \bibinfo {author} {\bibfnamefont {Z.}~\bibnamefont {Zhu}}, \bibinfo {author} {\bibfnamefont {Y.}~\bibnamefont {Liu}}, \bibinfo {author} {\bibfnamefont {Z.}~\bibnamefont {Zhang}}, \bibinfo {author} {\bibfnamefont {I.}~\bibnamefont {{\v{Z}}uti{\'c}}},\ and\ \bibinfo {author} {\bibfnamefont {T.}~\bibnamefont {Zhou}},\ }\bibfield  {title} {\bibinfo {title} {Antiferroelectric altermagnets: Antiferroelectricity alters magnets},\ }\href {https://doi.org/10.1103/PhysRevLett.134.106801} {\bibfield  {journal} {\bibinfo  {journal} {Phys. Rev. Lett.}\ }\textbf {\bibinfo {volume} {134}},\ \bibinfo {pages} {106801} (\bibinfo {year} {2025})}\BibitemShut {NoStop}%
\bibitem [{\citenamefont {Chen}\ \emph {et~al.}(2025)\citenamefont {Chen}, \citenamefont {Liu}, \citenamefont {Lu},\ and\ \citenamefont {Xie}}]{chen2025electrical}%
  \BibitemOpen
  \bibfield  {author} {\bibinfo {author} {\bibfnamefont {Y.}~\bibnamefont {Chen}}, \bibinfo {author} {\bibfnamefont {X.}~\bibnamefont {Liu}}, \bibinfo {author} {\bibfnamefont {H.-Z.}\ \bibnamefont {Lu}},\ and\ \bibinfo {author} {\bibfnamefont {X.}~\bibnamefont {Xie}},\ }\bibfield  {title} {\bibinfo {title} {Electrical switching of altermagnetism},\ }\href@noop {} {\bibfield  {journal} {\bibinfo  {journal} {Physical Review Letters}\ }\textbf {\bibinfo {volume} {135}},\ \bibinfo {pages} {016701} (\bibinfo {year} {2025})}\BibitemShut {NoStop}%
\bibitem [{\citenamefont {Gu}\ \emph {et~al.}(2025)\citenamefont {Gu}, \citenamefont {Liu}, \citenamefont {Zhu}, \citenamefont {Yananose}, \citenamefont {Chen}, \citenamefont {Hu}, \citenamefont {Stroppa},\ and\ \citenamefont {Liu}}]{gu2025ferroelectric}%
  \BibitemOpen
  \bibfield  {author} {\bibinfo {author} {\bibfnamefont {M.}~\bibnamefont {Gu}}, \bibinfo {author} {\bibfnamefont {Y.}~\bibnamefont {Liu}}, \bibinfo {author} {\bibfnamefont {H.}~\bibnamefont {Zhu}}, \bibinfo {author} {\bibfnamefont {K.}~\bibnamefont {Yananose}}, \bibinfo {author} {\bibfnamefont {X.}~\bibnamefont {Chen}}, \bibinfo {author} {\bibfnamefont {Y.}~\bibnamefont {Hu}}, \bibinfo {author} {\bibfnamefont {A.}~\bibnamefont {Stroppa}},\ and\ \bibinfo {author} {\bibfnamefont {Q.}~\bibnamefont {Liu}},\ }\bibfield  {title} {\bibinfo {title} {Ferroelectric switchable altermagnetism},\ }\href {https://doi.org/10.1103/PhysRevLett.134.106802} {\bibfield  {journal} {\bibinfo  {journal} {Phys. Rev. Lett.}\ }\textbf {\bibinfo {volume} {134}},\ \bibinfo {pages} {106802} (\bibinfo {year} {2025})}\BibitemShut {NoStop}%
\bibitem [{\citenamefont {{\v{S}}mejkal}(2024)}]{smejkal2024altermagnetic}%
  \BibitemOpen
  \bibfield  {author} {\bibinfo {author} {\bibfnamefont {L.}~\bibnamefont {{\v{S}}mejkal}},\ }\bibfield  {title} {\bibinfo {title} {Altermagnetic multiferroics and altermagnetoelectric effect},\ }\href@noop {} {\bibfield  {journal} {\bibinfo  {journal} {arXiv preprint arXiv:2411.19928}\ } (\bibinfo {year} {2024})}\BibitemShut {NoStop}%
\bibitem [{\citenamefont {Fukaya}\ \emph {et~al.}(2025)\citenamefont {Fukaya}, \citenamefont {Lu}, \citenamefont {Yada}, \citenamefont {Tanaka},\ and\ \citenamefont {Cayao}}]{Fukaya_2025}%
  \BibitemOpen
  \bibfield  {author} {\bibinfo {author} {\bibfnamefont {Y.}~\bibnamefont {Fukaya}}, \bibinfo {author} {\bibfnamefont {B.}~\bibnamefont {Lu}}, \bibinfo {author} {\bibfnamefont {K.}~\bibnamefont {Yada}}, \bibinfo {author} {\bibfnamefont {Y.}~\bibnamefont {Tanaka}},\ and\ \bibinfo {author} {\bibfnamefont {J.}~\bibnamefont {Cayao}},\ }\bibfield  {title} {\bibinfo {title} {Superconducting phenomena in systems with unconventional magnets},\ }\href {https://doi.org/10.1088/1361-648X/adf1cf} {\bibfield  {journal} {\bibinfo  {journal} {Journal of Physics: Condensed Matter}\ }\textbf {\bibinfo {volume} {37}},\ \bibinfo {pages} {313003} (\bibinfo {year} {2025})}\BibitemShut {NoStop}%
\bibitem [{\citenamefont {Hasan}\ and\ \citenamefont {Kane}(2010)}]{hasan2010colloquium}%
  \BibitemOpen
  \bibfield  {author} {\bibinfo {author} {\bibfnamefont {M.~Z.}\ \bibnamefont {Hasan}}\ and\ \bibinfo {author} {\bibfnamefont {C.~L.}\ \bibnamefont {Kane}},\ }\bibfield  {title} {\bibinfo {title} {Colloquium: topological insulators},\ }\href@noop {} {\bibfield  {journal} {\bibinfo  {journal} {Reviews of modern physics}\ }\textbf {\bibinfo {volume} {82}},\ \bibinfo {pages} {3045} (\bibinfo {year} {2010})}\BibitemShut {NoStop}%
\bibitem [{\citenamefont {Qi}\ and\ \citenamefont {Zhang}(2011)}]{qi2011topological}%
  \BibitemOpen
  \bibfield  {author} {\bibinfo {author} {\bibfnamefont {X.-L.}\ \bibnamefont {Qi}}\ and\ \bibinfo {author} {\bibfnamefont {S.-C.}\ \bibnamefont {Zhang}},\ }\bibfield  {title} {\bibinfo {title} {Topological insulators and superconductors},\ }\href {https://doi.org/10.1103/RevModPhys.83.1057} {\bibfield  {journal} {\bibinfo  {journal} {Rev. Mod. Phys.}\ }\textbf {\bibinfo {volume} {83}},\ \bibinfo {pages} {1057} (\bibinfo {year} {2011})}\BibitemShut {NoStop}%
\bibitem [{\citenamefont {Wieder}\ \emph {et~al.}(2022)\citenamefont {Wieder}, \citenamefont {Bradlyn}, \citenamefont {Cano}, \citenamefont {Wang}, \citenamefont {Vergniory}, \citenamefont {Elcoro}, \citenamefont {Soluyanov}, \citenamefont {Felser}, \citenamefont {Neupert}, \citenamefont {Regnault},\ and\ \citenamefont {Bernevig}}]{wieder2022topological}%
  \BibitemOpen
  \bibfield  {author} {\bibinfo {author} {\bibfnamefont {B.~J.}\ \bibnamefont {Wieder}}, \bibinfo {author} {\bibfnamefont {B.}~\bibnamefont {Bradlyn}}, \bibinfo {author} {\bibfnamefont {J.}~\bibnamefont {Cano}}, \bibinfo {author} {\bibfnamefont {Z.}~\bibnamefont {Wang}}, \bibinfo {author} {\bibfnamefont {M.~G.}\ \bibnamefont {Vergniory}}, \bibinfo {author} {\bibfnamefont {L.}~\bibnamefont {Elcoro}}, \bibinfo {author} {\bibfnamefont {A.~A.}\ \bibnamefont {Soluyanov}}, \bibinfo {author} {\bibfnamefont {C.}~\bibnamefont {Felser}}, \bibinfo {author} {\bibfnamefont {T.}~\bibnamefont {Neupert}}, \bibinfo {author} {\bibfnamefont {N.}~\bibnamefont {Regnault}},\ and\ \bibinfo {author} {\bibfnamefont {B.~A.}\ \bibnamefont {Bernevig}},\ }\bibfield  {title} {\bibinfo {title} {Topological materials discovery from crystal symmetry},\ }\href@noop {} {\bibfield  {journal} {\bibinfo  {journal} {Nature Reviews Materials}\ }\textbf {\bibinfo {volume} {7}},\ \bibinfo {pages} {196} (\bibinfo {year} {2022})}\BibitemShut {NoStop}%
\bibitem [{\citenamefont {Hsieh}\ \emph {et~al.}(2012)\citenamefont {Hsieh}, \citenamefont {Lin}, \citenamefont {Liu}, \citenamefont {Duan}, \citenamefont {Bansil},\ and\ \citenamefont {Fu}}]{hsieh2012topological}%
  \BibitemOpen
  \bibfield  {author} {\bibinfo {author} {\bibfnamefont {T.~H.}\ \bibnamefont {Hsieh}}, \bibinfo {author} {\bibfnamefont {H.}~\bibnamefont {Lin}}, \bibinfo {author} {\bibfnamefont {J.}~\bibnamefont {Liu}}, \bibinfo {author} {\bibfnamefont {W.}~\bibnamefont {Duan}}, \bibinfo {author} {\bibfnamefont {A.}~\bibnamefont {Bansil}},\ and\ \bibinfo {author} {\bibfnamefont {L.}~\bibnamefont {Fu}},\ }\bibfield  {title} {\bibinfo {title} {Topological crystalline insulators in the {SnTe} material class},\ }\href {https://doi.org/10.1038/ncomms1969} {\bibfield  {journal} {\bibinfo  {journal} {Nature communications}\ }\textbf {\bibinfo {volume} {3}},\ \bibinfo {pages} {982} (\bibinfo {year} {2012})}\BibitemShut {NoStop}%
\bibitem [{\citenamefont {Tanaka}\ \emph {et~al.}(2012)\citenamefont {Tanaka}, \citenamefont {Ren}, \citenamefont {Sato}, \citenamefont {Nakayama}, \citenamefont {Souma}, \citenamefont {Takahashi}, \citenamefont {Segawa},\ and\ \citenamefont {Ando}}]{tanaka2012experimental}%
  \BibitemOpen
  \bibfield  {author} {\bibinfo {author} {\bibfnamefont {Y.}~\bibnamefont {Tanaka}}, \bibinfo {author} {\bibfnamefont {Z.}~\bibnamefont {Ren}}, \bibinfo {author} {\bibfnamefont {T.}~\bibnamefont {Sato}}, \bibinfo {author} {\bibfnamefont {K.}~\bibnamefont {Nakayama}}, \bibinfo {author} {\bibfnamefont {S.}~\bibnamefont {Souma}}, \bibinfo {author} {\bibfnamefont {T.}~\bibnamefont {Takahashi}}, \bibinfo {author} {\bibfnamefont {K.}~\bibnamefont {Segawa}},\ and\ \bibinfo {author} {\bibfnamefont {Y.}~\bibnamefont {Ando}},\ }\bibfield  {title} {\bibinfo {title} {Experimental realization of a topological crystalline insulator in {SnTe}},\ }\href@noop {} {\bibfield  {journal} {\bibinfo  {journal} {Nature Physics}\ }\textbf {\bibinfo {volume} {8}},\ \bibinfo {pages} {800} (\bibinfo {year} {2012})}\BibitemShut {NoStop}%
\bibitem [{\citenamefont {{\v{S}}mejkal}\ \emph {et~al.}(2022{\natexlab{d}})\citenamefont {{\v{S}}mejkal}, \citenamefont {Sinova},\ and\ \citenamefont {Jungwirth}}]{vsmejkal2022emerging}%
  \BibitemOpen
  \bibfield  {author} {\bibinfo {author} {\bibfnamefont {L.}~\bibnamefont {{\v{S}}mejkal}}, \bibinfo {author} {\bibfnamefont {J.}~\bibnamefont {Sinova}},\ and\ \bibinfo {author} {\bibfnamefont {T.}~\bibnamefont {Jungwirth}},\ }\bibfield  {title} {\bibinfo {title} {Emerging research landscape of altermagnetism},\ }\href@noop {} {\bibfield  {journal} {\bibinfo  {journal} {Physical Review X}\ }\textbf {\bibinfo {volume} {12}},\ \bibinfo {pages} {040501} (\bibinfo {year} {2022}{\natexlab{d}})}\BibitemShut {NoStop}%
\bibitem [{\citenamefont {Lin}\ \emph {et~al.}(2024)\citenamefont {Lin}, \citenamefont {Chen}, \citenamefont {Lu}, \citenamefont {Liang}, \citenamefont {Feng}, \citenamefont {Yamagami}, \citenamefont {Osiecki}, \citenamefont {Leandersson}, \citenamefont {Thiagarajan}, \citenamefont {Liu} \emph {et~al.}}]{lin2024observation}%
  \BibitemOpen
  \bibfield  {author} {\bibinfo {author} {\bibfnamefont {Z.}~\bibnamefont {Lin}}, \bibinfo {author} {\bibfnamefont {D.}~\bibnamefont {Chen}}, \bibinfo {author} {\bibfnamefont {W.}~\bibnamefont {Lu}}, \bibinfo {author} {\bibfnamefont {X.}~\bibnamefont {Liang}}, \bibinfo {author} {\bibfnamefont {S.}~\bibnamefont {Feng}}, \bibinfo {author} {\bibfnamefont {K.}~\bibnamefont {Yamagami}}, \bibinfo {author} {\bibfnamefont {J.}~\bibnamefont {Osiecki}}, \bibinfo {author} {\bibfnamefont {M.}~\bibnamefont {Leandersson}}, \bibinfo {author} {\bibfnamefont {B.}~\bibnamefont {Thiagarajan}}, \bibinfo {author} {\bibfnamefont {J.}~\bibnamefont {Liu}}, \emph {et~al.},\ }\bibfield  {title} {\bibinfo {title} {Observation of giant spin splitting and d-wave spin texture in room temperature altermagnet {RuO$_2$}},\ }\href@noop {} {\bibfield  {journal} {\bibinfo  {journal} {arXiv preprint arXiv:2402.04995}\ } (\bibinfo {year} {2024})}\BibitemShut {NoStop}%
\bibitem [{\citenamefont {Zhang}\ \emph {et~al.}(2025)\citenamefont {Zhang}, \citenamefont {Jeong}, \citenamefont {Buiarelli}, \citenamefont {Lee}, \citenamefont {Guo}, \citenamefont {Wen}, \citenamefont {Li}, \citenamefont {Nair}, \citenamefont {Choi}, \citenamefont {Ren}, \citenamefont {Yue}, \citenamefont {Fedorov}, \citenamefont {Mo}, \citenamefont {Kono}, \citenamefont {Lee}, \citenamefont {Low}, \citenamefont {Birol}, \citenamefont {Fernandes}, \citenamefont {Radovic}, \citenamefont {Jalan},\ and\ \citenamefont {Yi}}]{zhang2025observationmirroroddmirrorevenspin}%
  \BibitemOpen
  \bibfield  {author} {\bibinfo {author} {\bibfnamefont {Y.}~\bibnamefont {Zhang}}, \bibinfo {author} {\bibfnamefont {S.~G.}\ \bibnamefont {Jeong}}, \bibinfo {author} {\bibfnamefont {L.}~\bibnamefont {Buiarelli}}, \bibinfo {author} {\bibfnamefont {S.}~\bibnamefont {Lee}}, \bibinfo {author} {\bibfnamefont {Y.}~\bibnamefont {Guo}}, \bibinfo {author} {\bibfnamefont {J.}~\bibnamefont {Wen}}, \bibinfo {author} {\bibfnamefont {H.}~\bibnamefont {Li}}, \bibinfo {author} {\bibfnamefont {S.}~\bibnamefont {Nair}}, \bibinfo {author} {\bibfnamefont {I.~H.}\ \bibnamefont {Choi}}, \bibinfo {author} {\bibfnamefont {Z.}~\bibnamefont {Ren}}, \bibinfo {author} {\bibfnamefont {Z.}~\bibnamefont {Yue}}, \bibinfo {author} {\bibfnamefont {A.}~\bibnamefont {Fedorov}}, \bibinfo {author} {\bibfnamefont {S.-K.}\ \bibnamefont {Mo}}, \bibinfo {author} {\bibfnamefont {J.}~\bibnamefont {Kono}}, \bibinfo {author} {\bibfnamefont {J.~S.}\ \bibnamefont {Lee}}, \bibinfo {author} {\bibfnamefont {T.}~\bibnamefont {Low}}, \bibinfo {author}
  {\bibfnamefont {T.}~\bibnamefont {Birol}}, \bibinfo {author} {\bibfnamefont {R.~M.}\ \bibnamefont {Fernandes}}, \bibinfo {author} {\bibfnamefont {M.}~\bibnamefont {Radovic}}, \bibinfo {author} {\bibfnamefont {B.}~\bibnamefont {Jalan}},\ and\ \bibinfo {author} {\bibfnamefont {M.}~\bibnamefont {Yi}},\ }\href {https://arxiv.org/abs/2509.16361} {\bibinfo {title} {Observation of mirror-odd and mirror-even spin texture in ultra-thin epitaxially-strained ruo2 films}} (\bibinfo {year} {2025}),\ \Eprint {https://arxiv.org/abs/2509.16361} {arXiv:2509.16361 [cond-mat.mtrl-sci]} \BibitemShut {NoStop}%
\bibitem [{\citenamefont {Feng}\ \emph {et~al.}(2022)\citenamefont {Feng}, \citenamefont {Zhou}, \citenamefont {{\v{S}}mejkal}, \citenamefont {Wu}, \citenamefont {Zhu}, \citenamefont {Guo}, \citenamefont {Gonz{\'a}lez-Hern{\'a}ndez}, \citenamefont {Wang}, \citenamefont {Yan}, \citenamefont {Qin} \emph {et~al.}}]{feng2022anomalous}%
  \BibitemOpen
  \bibfield  {author} {\bibinfo {author} {\bibfnamefont {Z.}~\bibnamefont {Feng}}, \bibinfo {author} {\bibfnamefont {X.}~\bibnamefont {Zhou}}, \bibinfo {author} {\bibfnamefont {L.}~\bibnamefont {{\v{S}}mejkal}}, \bibinfo {author} {\bibfnamefont {L.}~\bibnamefont {Wu}}, \bibinfo {author} {\bibfnamefont {Z.}~\bibnamefont {Zhu}}, \bibinfo {author} {\bibfnamefont {H.}~\bibnamefont {Guo}}, \bibinfo {author} {\bibfnamefont {R.}~\bibnamefont {Gonz{\'a}lez-Hern{\'a}ndez}}, \bibinfo {author} {\bibfnamefont {X.}~\bibnamefont {Wang}}, \bibinfo {author} {\bibfnamefont {H.}~\bibnamefont {Yan}}, \bibinfo {author} {\bibfnamefont {P.}~\bibnamefont {Qin}}, \emph {et~al.},\ }\bibfield  {title} {\bibinfo {title} {An anomalous hall effect in altermagnetic ruthenium dioxide},\ }\href@noop {} {\bibfield  {journal} {\bibinfo  {journal} {Nature Electronics}\ }\textbf {\bibinfo {volume} {5}},\ \bibinfo {pages} {735} (\bibinfo {year} {2022})}\BibitemShut {NoStop}%
\bibitem [{\citenamefont {Bai}\ \emph {et~al.}(2023)\citenamefont {Bai}, \citenamefont {Zhang}, \citenamefont {Zhou}, \citenamefont {Chen}, \citenamefont {Wan}, \citenamefont {Han}, \citenamefont {Zhu}, \citenamefont {Liang}, \citenamefont {Su}, \citenamefont {Han}, \citenamefont {Pan},\ and\ \citenamefont {Song}}]{PhysRevLett.130.216701}%
  \BibitemOpen
  \bibfield  {author} {\bibinfo {author} {\bibfnamefont {H.}~\bibnamefont {Bai}}, \bibinfo {author} {\bibfnamefont {Y.~C.}\ \bibnamefont {Zhang}}, \bibinfo {author} {\bibfnamefont {Y.~J.}\ \bibnamefont {Zhou}}, \bibinfo {author} {\bibfnamefont {P.}~\bibnamefont {Chen}}, \bibinfo {author} {\bibfnamefont {C.~H.}\ \bibnamefont {Wan}}, \bibinfo {author} {\bibfnamefont {L.}~\bibnamefont {Han}}, \bibinfo {author} {\bibfnamefont {W.~X.}\ \bibnamefont {Zhu}}, \bibinfo {author} {\bibfnamefont {S.~X.}\ \bibnamefont {Liang}}, \bibinfo {author} {\bibfnamefont {Y.~C.}\ \bibnamefont {Su}}, \bibinfo {author} {\bibfnamefont {X.~F.}\ \bibnamefont {Han}}, \bibinfo {author} {\bibfnamefont {F.}~\bibnamefont {Pan}},\ and\ \bibinfo {author} {\bibfnamefont {C.}~\bibnamefont {Song}},\ }\bibfield  {title} {\bibinfo {title} {Efficient spin-to-charge conversion via altermagnetic spin splitting effect in antiferromagnet ${\mathrm{ruo}}_{2}$},\ }\href {https://doi.org/10.1103/PhysRevLett.130.216701} {\bibfield  {journal} {\bibinfo
  {journal} {Phys. Rev. Lett.}\ }\textbf {\bibinfo {volume} {130}},\ \bibinfo {pages} {216701} (\bibinfo {year} {2023})}\BibitemShut {NoStop}%
\bibitem [{\citenamefont {Jeong}\ \emph {et~al.}(2025)\citenamefont {Jeong}, \citenamefont {Choi}, \citenamefont {Nair}, \citenamefont {Buiarelli}, \citenamefont {Pourbahari}, \citenamefont {Oh}, \citenamefont {Bassim}, \citenamefont {Hirai}, \citenamefont {Seo}, \citenamefont {Choi}, \citenamefont {Fernandes}, \citenamefont {Birol}, \citenamefont {Zhao}, \citenamefont {Lee},\ and\ \citenamefont {Jalan}}]{jeong2025altermagneticpolarmetallicphase}%
  \BibitemOpen
  \bibfield  {author} {\bibinfo {author} {\bibfnamefont {S.~G.}\ \bibnamefont {Jeong}}, \bibinfo {author} {\bibfnamefont {I.~H.}\ \bibnamefont {Choi}}, \bibinfo {author} {\bibfnamefont {S.}~\bibnamefont {Nair}}, \bibinfo {author} {\bibfnamefont {L.}~\bibnamefont {Buiarelli}}, \bibinfo {author} {\bibfnamefont {B.}~\bibnamefont {Pourbahari}}, \bibinfo {author} {\bibfnamefont {J.~Y.}\ \bibnamefont {Oh}}, \bibinfo {author} {\bibfnamefont {N.}~\bibnamefont {Bassim}}, \bibinfo {author} {\bibfnamefont {D.}~\bibnamefont {Hirai}}, \bibinfo {author} {\bibfnamefont {A.}~\bibnamefont {Seo}}, \bibinfo {author} {\bibfnamefont {W.~S.}\ \bibnamefont {Choi}}, \bibinfo {author} {\bibfnamefont {R.~M.}\ \bibnamefont {Fernandes}}, \bibinfo {author} {\bibfnamefont {T.}~\bibnamefont {Birol}}, \bibinfo {author} {\bibfnamefont {L.}~\bibnamefont {Zhao}}, \bibinfo {author} {\bibfnamefont {J.~S.}\ \bibnamefont {Lee}},\ and\ \bibinfo {author} {\bibfnamefont {B.}~\bibnamefont {Jalan}},\ }\href {https://arxiv.org/abs/2405.05838} {\bibinfo
  {title} {Altermagnetic polar metallic phase in ultra-thin epitaxially-strained ruo2 films}} (\bibinfo {year} {2025}),\ \Eprint {https://arxiv.org/abs/2405.05838} {arXiv:2405.05838 [cond-mat.mtrl-sci]} \BibitemShut {NoStop}%
\bibitem [{\citenamefont {Fedchenko}\ \emph {et~al.}(2024)\citenamefont {Fedchenko}, \citenamefont {Min{\'a}r}, \citenamefont {Akashdeep}, \citenamefont {D’Souza}, \citenamefont {Vasilyev}, \citenamefont {Tkach}, \citenamefont {Odenbreit}, \citenamefont {Nguyen}, \citenamefont {Kutnyakhov}, \citenamefont {Wind} \emph {et~al.}}]{fedchenko2024observation}%
  \BibitemOpen
  \bibfield  {author} {\bibinfo {author} {\bibfnamefont {O.}~\bibnamefont {Fedchenko}}, \bibinfo {author} {\bibfnamefont {J.}~\bibnamefont {Min{\'a}r}}, \bibinfo {author} {\bibfnamefont {A.}~\bibnamefont {Akashdeep}}, \bibinfo {author} {\bibfnamefont {S.~W.}\ \bibnamefont {D’Souza}}, \bibinfo {author} {\bibfnamefont {D.}~\bibnamefont {Vasilyev}}, \bibinfo {author} {\bibfnamefont {O.}~\bibnamefont {Tkach}}, \bibinfo {author} {\bibfnamefont {L.}~\bibnamefont {Odenbreit}}, \bibinfo {author} {\bibfnamefont {Q.}~\bibnamefont {Nguyen}}, \bibinfo {author} {\bibfnamefont {D.}~\bibnamefont {Kutnyakhov}}, \bibinfo {author} {\bibfnamefont {N.}~\bibnamefont {Wind}}, \emph {et~al.},\ }\bibfield  {title} {\bibinfo {title} {Observation of time-reversal symmetry breaking in the band structure of altermagnetic {RuO$_2$}},\ }\href@noop {} {\bibfield  {journal} {\bibinfo  {journal} {Science advances}\ }\textbf {\bibinfo {volume} {10}},\ \bibinfo {pages} {eadj4883} (\bibinfo {year} {2024})}\BibitemShut {NoStop}%
\bibitem [{\citenamefont {Liu}\ \emph {et~al.}(2024)\citenamefont {Liu}, \citenamefont {Zhan}, \citenamefont {Li}, \citenamefont {Liu}, \citenamefont {Cheng}, \citenamefont {Shi}, \citenamefont {Deng}, \citenamefont {Zhang}, \citenamefont {Li}, \citenamefont {Ding} \emph {et~al.}}]{liu2024absence}%
  \BibitemOpen
  \bibfield  {author} {\bibinfo {author} {\bibfnamefont {J.}~\bibnamefont {Liu}}, \bibinfo {author} {\bibfnamefont {J.}~\bibnamefont {Zhan}}, \bibinfo {author} {\bibfnamefont {T.}~\bibnamefont {Li}}, \bibinfo {author} {\bibfnamefont {J.}~\bibnamefont {Liu}}, \bibinfo {author} {\bibfnamefont {S.}~\bibnamefont {Cheng}}, \bibinfo {author} {\bibfnamefont {Y.}~\bibnamefont {Shi}}, \bibinfo {author} {\bibfnamefont {L.}~\bibnamefont {Deng}}, \bibinfo {author} {\bibfnamefont {M.}~\bibnamefont {Zhang}}, \bibinfo {author} {\bibfnamefont {C.}~\bibnamefont {Li}}, \bibinfo {author} {\bibfnamefont {J.}~\bibnamefont {Ding}}, \emph {et~al.},\ }\bibfield  {title} {\bibinfo {title} {Absence of altermagnetic spin splitting character in rutile oxide {RuO$_2$}},\ }\href@noop {} {\bibfield  {journal} {\bibinfo  {journal} {Physical Review Letters}\ }\textbf {\bibinfo {volume} {133}},\ \bibinfo {pages} {176401} (\bibinfo {year} {2024})}\BibitemShut {NoStop}%
\bibitem [{\citenamefont {Ke{\ss}ler}\ \emph {et~al.}(2024)\citenamefont {Ke{\ss}ler}, \citenamefont {Garcia-Gassull}, \citenamefont {Suter}, \citenamefont {Prokscha}, \citenamefont {Salman}, \citenamefont {Khalyavin}, \citenamefont {Manuel}, \citenamefont {Orlandi}, \citenamefont {Mazin}, \citenamefont {Valent{\'\i}} \emph {et~al.}}]{kessler2024absence}%
  \BibitemOpen
  \bibfield  {author} {\bibinfo {author} {\bibfnamefont {P.}~\bibnamefont {Ke{\ss}ler}}, \bibinfo {author} {\bibfnamefont {L.}~\bibnamefont {Garcia-Gassull}}, \bibinfo {author} {\bibfnamefont {A.}~\bibnamefont {Suter}}, \bibinfo {author} {\bibfnamefont {T.}~\bibnamefont {Prokscha}}, \bibinfo {author} {\bibfnamefont {Z.}~\bibnamefont {Salman}}, \bibinfo {author} {\bibfnamefont {D.}~\bibnamefont {Khalyavin}}, \bibinfo {author} {\bibfnamefont {P.}~\bibnamefont {Manuel}}, \bibinfo {author} {\bibfnamefont {F.}~\bibnamefont {Orlandi}}, \bibinfo {author} {\bibfnamefont {I.~I.}\ \bibnamefont {Mazin}}, \bibinfo {author} {\bibfnamefont {R.}~\bibnamefont {Valent{\'\i}}}, \emph {et~al.},\ }\bibfield  {title} {\bibinfo {title} {Absence of magnetic order in ruo2: insights from $\mu$ sr spectroscopy and neutron diffraction},\ }\href@noop {} {\bibfield  {journal} {\bibinfo  {journal} {npj Spintronics}\ }\textbf {\bibinfo {volume} {2}},\ \bibinfo {pages} {50} (\bibinfo {year} {2024})}\BibitemShut {NoStop}%
\bibitem [{\citenamefont {Smolyanyuk}\ \emph {et~al.}(2024)\citenamefont {Smolyanyuk}, \citenamefont {Mazin}, \citenamefont {Garcia-Gassull},\ and\ \citenamefont {Valent\'{\i}}}]{PhysRevB.109.134424}%
  \BibitemOpen
  \bibfield  {author} {\bibinfo {author} {\bibfnamefont {A.}~\bibnamefont {Smolyanyuk}}, \bibinfo {author} {\bibfnamefont {I.~I.}\ \bibnamefont {Mazin}}, \bibinfo {author} {\bibfnamefont {L.}~\bibnamefont {Garcia-Gassull}},\ and\ \bibinfo {author} {\bibfnamefont {R.}~\bibnamefont {Valent\'{\i}}},\ }\bibfield  {title} {\bibinfo {title} {Fragility of the magnetic order in the prototypical altermagnet ${\mathrm{ruo}}_{2}$},\ }\href {https://doi.org/10.1103/PhysRevB.109.134424} {\bibfield  {journal} {\bibinfo  {journal} {Phys. Rev. B}\ }\textbf {\bibinfo {volume} {109}},\ \bibinfo {pages} {134424} (\bibinfo {year} {2024})}\BibitemShut {NoStop}%
\bibitem [{\citenamefont {Qian}\ \emph {et~al.}(2025)\citenamefont {Qian}, \citenamefont {Rutherford}, \citenamefont {Choi}, \citenamefont {Zhou}, \citenamefont {Maiorov}, \citenamefont {Lee},\ and\ \citenamefont {Mizzi}}]{qian2025determiningnaturemagnetismaltermagnetic}%
  \BibitemOpen
  \bibfield  {author} {\bibinfo {author} {\bibfnamefont {T.}~\bibnamefont {Qian}}, \bibinfo {author} {\bibfnamefont {A.}~\bibnamefont {Rutherford}}, \bibinfo {author} {\bibfnamefont {E.~S.}\ \bibnamefont {Choi}}, \bibinfo {author} {\bibfnamefont {H.}~\bibnamefont {Zhou}}, \bibinfo {author} {\bibfnamefont {B.}~\bibnamefont {Maiorov}}, \bibinfo {author} {\bibfnamefont {M.}~\bibnamefont {Lee}},\ and\ \bibinfo {author} {\bibfnamefont {C.~A.}\ \bibnamefont {Mizzi}},\ }\href {https://arxiv.org/abs/2504.21138} {\bibinfo {title} {Determining the nature of magnetism in altermagnetic candidate ruo$_2$}} (\bibinfo {year} {2025}),\ \Eprint {https://arxiv.org/abs/2504.21138} {arXiv:2504.21138 [cond-mat.mtrl-sci]} \BibitemShut {NoStop}%
\bibitem [{\citenamefont {Hiraishi}\ \emph {et~al.}(2024)\citenamefont {Hiraishi}, \citenamefont {Okabe}, \citenamefont {Koda}, \citenamefont {Kadono}, \citenamefont {Muroi}, \citenamefont {Hirai},\ and\ \citenamefont {Hiroi}}]{PhysRevLett.132.166702}%
  \BibitemOpen
  \bibfield  {author} {\bibinfo {author} {\bibfnamefont {M.}~\bibnamefont {Hiraishi}}, \bibinfo {author} {\bibfnamefont {H.}~\bibnamefont {Okabe}}, \bibinfo {author} {\bibfnamefont {A.}~\bibnamefont {Koda}}, \bibinfo {author} {\bibfnamefont {R.}~\bibnamefont {Kadono}}, \bibinfo {author} {\bibfnamefont {T.}~\bibnamefont {Muroi}}, \bibinfo {author} {\bibfnamefont {D.}~\bibnamefont {Hirai}},\ and\ \bibinfo {author} {\bibfnamefont {Z.}~\bibnamefont {Hiroi}},\ }\bibfield  {title} {\bibinfo {title} {Nonmagnetic ground state in ${\mathrm{ruo}}_{2}$ revealed by muon spin rotation},\ }\href {https://doi.org/10.1103/PhysRevLett.132.166702} {\bibfield  {journal} {\bibinfo  {journal} {Phys. Rev. Lett.}\ }\textbf {\bibinfo {volume} {132}},\ \bibinfo {pages} {166702} (\bibinfo {year} {2024})}\BibitemShut {NoStop}%
\bibitem [{\citenamefont {Chang}\ \emph {et~al.}(2016)\citenamefont {Chang}, \citenamefont {Liu}, \citenamefont {Lin}, \citenamefont {Wang}, \citenamefont {Zhao}, \citenamefont {Zhang}, \citenamefont {Jin}, \citenamefont {Zhong}, \citenamefont {Hu}, \citenamefont {Duan} \emph {et~al.}}]{chang2016discovery}%
  \BibitemOpen
  \bibfield  {author} {\bibinfo {author} {\bibfnamefont {K.}~\bibnamefont {Chang}}, \bibinfo {author} {\bibfnamefont {J.}~\bibnamefont {Liu}}, \bibinfo {author} {\bibfnamefont {H.}~\bibnamefont {Lin}}, \bibinfo {author} {\bibfnamefont {N.}~\bibnamefont {Wang}}, \bibinfo {author} {\bibfnamefont {K.}~\bibnamefont {Zhao}}, \bibinfo {author} {\bibfnamefont {A.}~\bibnamefont {Zhang}}, \bibinfo {author} {\bibfnamefont {F.}~\bibnamefont {Jin}}, \bibinfo {author} {\bibfnamefont {Y.}~\bibnamefont {Zhong}}, \bibinfo {author} {\bibfnamefont {X.}~\bibnamefont {Hu}}, \bibinfo {author} {\bibfnamefont {W.}~\bibnamefont {Duan}}, \emph {et~al.},\ }\bibfield  {title} {\bibinfo {title} {Discovery of robust in-plane ferroelectricity in atomic-thick {SnTe}},\ }\href@noop {} {\bibfield  {journal} {\bibinfo  {journal} {Science}\ }\textbf {\bibinfo {volume} {353}},\ \bibinfo {pages} {274} (\bibinfo {year} {2016})}\BibitemShut {NoStop}%
\bibitem [{\citenamefont {Shen}\ \emph {et~al.}(2014)\citenamefont {Shen}, \citenamefont {Jung}, \citenamefont {Disa}, \citenamefont {Walker}, \citenamefont {Ahn},\ and\ \citenamefont {Cha}}]{shen2014synthesis}%
  \BibitemOpen
  \bibfield  {author} {\bibinfo {author} {\bibfnamefont {J.}~\bibnamefont {Shen}}, \bibinfo {author} {\bibfnamefont {Y.}~\bibnamefont {Jung}}, \bibinfo {author} {\bibfnamefont {A.~S.}\ \bibnamefont {Disa}}, \bibinfo {author} {\bibfnamefont {F.~J.}\ \bibnamefont {Walker}}, \bibinfo {author} {\bibfnamefont {C.~H.}\ \bibnamefont {Ahn}},\ and\ \bibinfo {author} {\bibfnamefont {J.~J.}\ \bibnamefont {Cha}},\ }\bibfield  {title} {\bibinfo {title} {Synthesis of {SnTe} nanoplates with $\{$100$\}$ and $\{$111$\}$ surfaces},\ }\href@noop {} {\bibfield  {journal} {\bibinfo  {journal} {Nano letters}\ }\textbf {\bibinfo {volume} {14}},\ \bibinfo {pages} {4183} (\bibinfo {year} {2014})}\BibitemShut {NoStop}%
\bibitem [{\citenamefont {Liu}\ \emph {et~al.}(2013)\citenamefont {Liu}, \citenamefont {Duan},\ and\ \citenamefont {Fu}}]{liu2013two}%
  \BibitemOpen
  \bibfield  {author} {\bibinfo {author} {\bibfnamefont {J.}~\bibnamefont {Liu}}, \bibinfo {author} {\bibfnamefont {W.}~\bibnamefont {Duan}},\ and\ \bibinfo {author} {\bibfnamefont {L.}~\bibnamefont {Fu}},\ }\bibfield  {title} {\bibinfo {title} {Two types of surface states in topological crystalline insulators},\ }\href {https://doi.org/10.1103/PhysRevB.88.241303} {\bibfield  {journal} {\bibinfo  {journal} {Phys. Rev. B}\ }\textbf {\bibinfo {volume} {88}},\ \bibinfo {pages} {241303} (\bibinfo {year} {2013})}\BibitemShut {NoStop}%
\bibitem [{\citenamefont {Wang}\ \emph {et~al.}(2013)\citenamefont {Wang}, \citenamefont {Tsai}, \citenamefont {Lin}, \citenamefont {Xu}, \citenamefont {Neupane}, \citenamefont {Hasan},\ and\ \citenamefont {Bansil}}]{wang2013nontrivial}%
  \BibitemOpen
  \bibfield  {author} {\bibinfo {author} {\bibfnamefont {Y.~J.}\ \bibnamefont {Wang}}, \bibinfo {author} {\bibfnamefont {W.-F.}\ \bibnamefont {Tsai}}, \bibinfo {author} {\bibfnamefont {H.}~\bibnamefont {Lin}}, \bibinfo {author} {\bibfnamefont {S.-Y.}\ \bibnamefont {Xu}}, \bibinfo {author} {\bibfnamefont {M.}~\bibnamefont {Neupane}}, \bibinfo {author} {\bibfnamefont {M.~Z.}\ \bibnamefont {Hasan}},\ and\ \bibinfo {author} {\bibfnamefont {A.}~\bibnamefont {Bansil}},\ }\bibfield  {title} {\bibinfo {title} {Nontrivial spin texture of the coaxial dirac cones on the surface of topological crystalline insulator {SnTe}},\ }\href {https://doi.org/10.1103/PhysRevB.87.235317} {\bibfield  {journal} {\bibinfo  {journal} {Phys. Rev. B}\ }\textbf {\bibinfo {volume} {87}},\ \bibinfo {pages} {235317} (\bibinfo {year} {2013})}\BibitemShut {NoStop}%
\bibitem [{\citenamefont {Kresse}\ and\ \citenamefont {Joubert}(1999)}]{kresse1999ultrasoft}%
  \BibitemOpen
  \bibfield  {author} {\bibinfo {author} {\bibfnamefont {G.}~\bibnamefont {Kresse}}\ and\ \bibinfo {author} {\bibfnamefont {D.}~\bibnamefont {Joubert}},\ }\bibfield  {title} {\bibinfo {title} {From ultrasoft pseudopotentials to the projector augmented-wave method},\ }\href@noop {} {\bibfield  {journal} {\bibinfo  {journal} {Physical Review B}\ }\textbf {\bibinfo {volume} {59}},\ \bibinfo {pages} {1758} (\bibinfo {year} {1999})}\BibitemShut {NoStop}%
\bibitem [{\citenamefont {Kresse}\ and\ \citenamefont {Hafner}(1993)}]{kresse1993ab}%
  \BibitemOpen
  \bibfield  {author} {\bibinfo {author} {\bibfnamefont {G.}~\bibnamefont {Kresse}}\ and\ \bibinfo {author} {\bibfnamefont {J.}~\bibnamefont {Hafner}},\ }\bibfield  {title} {\bibinfo {title} {Ab initio molecular dynamics for liquid metals},\ }\href@noop {} {\bibfield  {journal} {\bibinfo  {journal} {Physical Review B}\ }\textbf {\bibinfo {volume} {47}},\ \bibinfo {pages} {558} (\bibinfo {year} {1993})}\BibitemShut {NoStop}%
\bibitem [{\citenamefont {Kresse}\ and\ \citenamefont {Furthmüller}(1996{\natexlab{a}})}]{kresse1996efficiency}%
  \BibitemOpen
  \bibfield  {author} {\bibinfo {author} {\bibfnamefont {G.}~\bibnamefont {Kresse}}\ and\ \bibinfo {author} {\bibfnamefont {J.}~\bibnamefont {Furthmüller}},\ }\bibfield  {title} {\bibinfo {title} {Efficiency of ab-initio total energy calculations for metals and semiconductors using a plane-wave basis set},\ }\href@noop {} {\bibfield  {journal} {\bibinfo  {journal} {Computational Materials Science}\ }\textbf {\bibinfo {volume} {6}},\ \bibinfo {pages} {15} (\bibinfo {year} {1996}{\natexlab{a}})}\BibitemShut {NoStop}%
\bibitem [{\citenamefont {Kresse}\ and\ \citenamefont {Furthmüller}(1996{\natexlab{b}})}]{kresse1996efficient}%
  \BibitemOpen
  \bibfield  {author} {\bibinfo {author} {\bibfnamefont {G.}~\bibnamefont {Kresse}}\ and\ \bibinfo {author} {\bibfnamefont {J.}~\bibnamefont {Furthmüller}},\ }\bibfield  {title} {\bibinfo {title} {Efficient iterative schemes for ab initio total-energy calculations using a plane-wave basis set},\ }\href@noop {} {\bibfield  {journal} {\bibinfo  {journal} {Physical Review B}\ }\textbf {\bibinfo {volume} {54}},\ \bibinfo {pages} {11169} (\bibinfo {year} {1996}{\natexlab{b}})}\BibitemShut {NoStop}%
\bibitem [{\citenamefont {Perdew}\ \emph {et~al.}(1996)\citenamefont {Perdew}, \citenamefont {Burke},\ and\ \citenamefont {Ernzerhof}}]{perdew1996generalized}%
  \BibitemOpen
  \bibfield  {author} {\bibinfo {author} {\bibfnamefont {J.~P.}\ \bibnamefont {Perdew}}, \bibinfo {author} {\bibfnamefont {K.}~\bibnamefont {Burke}},\ and\ \bibinfo {author} {\bibfnamefont {M.}~\bibnamefont {Ernzerhof}},\ }\bibfield  {title} {\bibinfo {title} {Generalized gradient approximation made simple},\ }\href@noop {} {\bibfield  {journal} {\bibinfo  {journal} {Physical Review Letters}\ }\textbf {\bibinfo {volume} {77}},\ \bibinfo {pages} {3865} (\bibinfo {year} {1996})}\BibitemShut {NoStop}%
\bibitem [{\citenamefont {Cococcioni}\ and\ \citenamefont {de~Gironcoli}(2005)}]{cococcioni2005linear}%
  \BibitemOpen
  \bibfield  {author} {\bibinfo {author} {\bibfnamefont {M.}~\bibnamefont {Cococcioni}}\ and\ \bibinfo {author} {\bibfnamefont {S.}~\bibnamefont {de~Gironcoli}},\ }\bibfield  {title} {\bibinfo {title} {Linear response approach to the calculation of the effective interaction parameters in the $\mathrm{LDA}+\mathrm{U}$ method},\ }\href {https://doi.org/10.1103/PhysRevB.71.035105} {\bibfield  {journal} {\bibinfo  {journal} {Phys. Rev. B}\ }\textbf {\bibinfo {volume} {71}},\ \bibinfo {pages} {035105} (\bibinfo {year} {2005})}\BibitemShut {NoStop}%
\bibitem [{\citenamefont {Zheng}(2025)}]{zheng2025vaspbandunfolding}%
  \BibitemOpen
  \bibfield  {author} {\bibinfo {author} {\bibfnamefont {Q.}~\bibnamefont {Zheng}},\ }\href@noop {} {\bibinfo {title} {Vasp band unfolding}},\ \bibinfo {howpublished} {\url{https://github.com/QijingZheng/VaspBandUnfolding}} (\bibinfo {year} {2025})\BibitemShut {NoStop}%
\bibitem [{\citenamefont {Yang}\ \emph {et~al.}(2020)\citenamefont {Yang}, \citenamefont {Wu},\ and\ \citenamefont {Marom}}]{yang2020topological}%
  \BibitemOpen
  \bibfield  {author} {\bibinfo {author} {\bibfnamefont {S.}~\bibnamefont {Yang}}, \bibinfo {author} {\bibfnamefont {C.}~\bibnamefont {Wu}},\ and\ \bibinfo {author} {\bibfnamefont {N.}~\bibnamefont {Marom}},\ }\bibfield  {title} {\bibinfo {title} {Topological properties of {SnSe/EuS} and {SnTe/CaTe} interfaces},\ }\href@noop {} {\bibfield  {journal} {\bibinfo  {journal} {Physical Review Materials}\ }\textbf {\bibinfo {volume} {4}},\ \bibinfo {pages} {034203} (\bibinfo {year} {2020})}\BibitemShut {NoStop}%
\bibitem [{\citenamefont {Roig}\ \emph {et~al.}(2024)\citenamefont {Roig}, \citenamefont {Kreisel}, \citenamefont {Yu}, \citenamefont {Andersen},\ and\ \citenamefont {Agterberg}}]{roig2024minimal}%
  \BibitemOpen
  \bibfield  {author} {\bibinfo {author} {\bibfnamefont {M.}~\bibnamefont {Roig}}, \bibinfo {author} {\bibfnamefont {A.}~\bibnamefont {Kreisel}}, \bibinfo {author} {\bibfnamefont {Y.}~\bibnamefont {Yu}}, \bibinfo {author} {\bibfnamefont {B.~M.}\ \bibnamefont {Andersen}},\ and\ \bibinfo {author} {\bibfnamefont {D.~F.}\ \bibnamefont {Agterberg}},\ }\bibfield  {title} {\bibinfo {title} {Minimal models for altermagnetism},\ }\href@noop {} {\bibfield  {journal} {\bibinfo  {journal} {Physical Review B}\ }\textbf {\bibinfo {volume} {110}},\ \bibinfo {pages} {144412} (\bibinfo {year} {2024})}\BibitemShut {NoStop}%
\bibitem [{sup()}]{supplemental2025}%
  \BibitemOpen
  \href@noop {} {\bibinfo {title} {See supplemental material at}},\ \bibinfo {note} {uRL will be inserted by publisher}\BibitemShut {NoStop}%
\bibitem [{\citenamefont {Varnava}\ and\ \citenamefont {Vanderbilt}(2018)}]{varnava2018surfaces}%
  \BibitemOpen
  \bibfield  {author} {\bibinfo {author} {\bibfnamefont {N.}~\bibnamefont {Varnava}}\ and\ \bibinfo {author} {\bibfnamefont {D.}~\bibnamefont {Vanderbilt}},\ }\bibfield  {title} {\bibinfo {title} {Surfaces of axion insulators},\ }\href@noop {} {\bibfield  {journal} {\bibinfo  {journal} {Physical Review B}\ }\textbf {\bibinfo {volume} {98}},\ \bibinfo {pages} {245117} (\bibinfo {year} {2018})}\BibitemShut {NoStop}%
\bibitem [{\citenamefont {Sato}\ and\ \citenamefont {Ando}(2017)}]{sato2017topological}%
  \BibitemOpen
  \bibfield  {author} {\bibinfo {author} {\bibfnamefont {M.}~\bibnamefont {Sato}}\ and\ \bibinfo {author} {\bibfnamefont {Y.}~\bibnamefont {Ando}},\ }\bibfield  {title} {\bibinfo {title} {Topological superconductors: a review},\ }\href@noop {} {\bibfield  {journal} {\bibinfo  {journal} {Reports on Progress in Physics}\ }\textbf {\bibinfo {volume} {80}},\ \bibinfo {pages} {076501} (\bibinfo {year} {2017})}\BibitemShut {NoStop}%
\bibitem [{\citenamefont {Chiu}\ \emph {et~al.}(2016)\citenamefont {Chiu}, \citenamefont {Teo}, \citenamefont {Schnyder},\ and\ \citenamefont {Ryu}}]{chiu2016classification}%
  \BibitemOpen
  \bibfield  {author} {\bibinfo {author} {\bibfnamefont {C.-K.}\ \bibnamefont {Chiu}}, \bibinfo {author} {\bibfnamefont {J.~C.}\ \bibnamefont {Teo}}, \bibinfo {author} {\bibfnamefont {A.~P.}\ \bibnamefont {Schnyder}},\ and\ \bibinfo {author} {\bibfnamefont {S.}~\bibnamefont {Ryu}},\ }\bibfield  {title} {\bibinfo {title} {Classification of topological quantum matter with symmetries},\ }\href@noop {} {\bibfield  {journal} {\bibinfo  {journal} {Reviews of Modern Physics}\ }\textbf {\bibinfo {volume} {88}},\ \bibinfo {pages} {035005} (\bibinfo {year} {2016})}\BibitemShut {NoStop}%
\bibitem [{\citenamefont {Nayak}\ \emph {et~al.}(2008)\citenamefont {Nayak}, \citenamefont {Simon}, \citenamefont {Stern}, \citenamefont {Freedman},\ and\ \citenamefont {Das~Sarma}}]{nayak2008non}%
  \BibitemOpen
  \bibfield  {author} {\bibinfo {author} {\bibfnamefont {C.}~\bibnamefont {Nayak}}, \bibinfo {author} {\bibfnamefont {S.~H.}\ \bibnamefont {Simon}}, \bibinfo {author} {\bibfnamefont {A.}~\bibnamefont {Stern}}, \bibinfo {author} {\bibfnamefont {M.}~\bibnamefont {Freedman}},\ and\ \bibinfo {author} {\bibfnamefont {S.}~\bibnamefont {Das~Sarma}},\ }\bibfield  {title} {\bibinfo {title} {Non-abelian anyons and topological quantum computation},\ }\href@noop {} {\bibfield  {journal} {\bibinfo  {journal} {Reviews of Modern Physics}\ }\textbf {\bibinfo {volume} {80}},\ \bibinfo {pages} {1083} (\bibinfo {year} {2008})}\BibitemShut {NoStop}%
\bibitem [{\citenamefont {Lutchyn}\ \emph {et~al.}(2010)\citenamefont {Lutchyn}, \citenamefont {Sau},\ and\ \citenamefont {Das~Sarma}}]{lutchyn2010majorana}%
  \BibitemOpen
  \bibfield  {author} {\bibinfo {author} {\bibfnamefont {R.~M.}\ \bibnamefont {Lutchyn}}, \bibinfo {author} {\bibfnamefont {J.~D.}\ \bibnamefont {Sau}},\ and\ \bibinfo {author} {\bibfnamefont {S.}~\bibnamefont {Das~Sarma}},\ }\bibfield  {title} {\bibinfo {title} {Majorana fermions and a topological phase transition in semiconductor-superconductor heterostructures},\ }\href@noop {} {\bibfield  {journal} {\bibinfo  {journal} {Physical Review Letters}\ }\textbf {\bibinfo {volume} {105}},\ \bibinfo {pages} {077001} (\bibinfo {year} {2010})}\BibitemShut {NoStop}%
\bibitem [{\citenamefont {Nadj-Perge}\ \emph {et~al.}(2013)\citenamefont {Nadj-Perge}, \citenamefont {Drozdov}, \citenamefont {Bernevig},\ and\ \citenamefont {Yazdani}}]{nadj2013proposal}%
  \BibitemOpen
  \bibfield  {author} {\bibinfo {author} {\bibfnamefont {S.}~\bibnamefont {Nadj-Perge}}, \bibinfo {author} {\bibfnamefont {I.~K.}\ \bibnamefont {Drozdov}}, \bibinfo {author} {\bibfnamefont {B.~A.}\ \bibnamefont {Bernevig}},\ and\ \bibinfo {author} {\bibfnamefont {A.}~\bibnamefont {Yazdani}},\ }\bibfield  {title} {\bibinfo {title} {Proposal for realizing majorana fermions in chains of magnetic atoms on a superconductor},\ }\href@noop {} {\bibfield  {journal} {\bibinfo  {journal} {Physical Review B}\ }\textbf {\bibinfo {volume} {88}},\ \bibinfo {pages} {020407} (\bibinfo {year} {2013})}\BibitemShut {NoStop}%
\bibitem [{\citenamefont {Nadj-Perge}\ \emph {et~al.}(2014)\citenamefont {Nadj-Perge}, \citenamefont {Drozdov}, \citenamefont {Li}, \citenamefont {Chen}, \citenamefont {Jeon}, \citenamefont {Seo}, \citenamefont {MacDonald}, \citenamefont {Bernevig},\ and\ \citenamefont {Yazdani}}]{nadj2014observation}%
  \BibitemOpen
  \bibfield  {author} {\bibinfo {author} {\bibfnamefont {S.}~\bibnamefont {Nadj-Perge}}, \bibinfo {author} {\bibfnamefont {I.~K.}\ \bibnamefont {Drozdov}}, \bibinfo {author} {\bibfnamefont {J.}~\bibnamefont {Li}}, \bibinfo {author} {\bibfnamefont {H.}~\bibnamefont {Chen}}, \bibinfo {author} {\bibfnamefont {S.}~\bibnamefont {Jeon}}, \bibinfo {author} {\bibfnamefont {J.}~\bibnamefont {Seo}}, \bibinfo {author} {\bibfnamefont {A.~H.}\ \bibnamefont {MacDonald}}, \bibinfo {author} {\bibfnamefont {B.~A.}\ \bibnamefont {Bernevig}},\ and\ \bibinfo {author} {\bibfnamefont {A.}~\bibnamefont {Yazdani}},\ }\bibfield  {title} {\bibinfo {title} {Observation of majorana fermions in ferromagnetic atomic chains on a superconductor},\ }\href@noop {} {\bibfield  {journal} {\bibinfo  {journal} {Science}\ }\textbf {\bibinfo {volume} {346}},\ \bibinfo {pages} {602} (\bibinfo {year} {2014})}\BibitemShut {NoStop}%
\bibitem [{\citenamefont {Qi}\ \emph {et~al.}(2010)\citenamefont {Qi}, \citenamefont {Hughes},\ and\ \citenamefont {Zhang}}]{qi2010chiral}%
  \BibitemOpen
  \bibfield  {author} {\bibinfo {author} {\bibfnamefont {X.-L.}\ \bibnamefont {Qi}}, \bibinfo {author} {\bibfnamefont {T.~L.}\ \bibnamefont {Hughes}},\ and\ \bibinfo {author} {\bibfnamefont {S.-C.}\ \bibnamefont {Zhang}},\ }\bibfield  {title} {\bibinfo {title} {Chiral topological superconductor from the quantum hall state},\ }\href@noop {} {\bibfield  {journal} {\bibinfo  {journal} {Physical Review B}\ }\textbf {\bibinfo {volume} {82}},\ \bibinfo {pages} {184516} (\bibinfo {year} {2010})}\BibitemShut {NoStop}%
\bibitem [{\citenamefont {Wang}\ \emph {et~al.}(2015)\citenamefont {Wang}, \citenamefont {Zhou}, \citenamefont {Lian},\ and\ \citenamefont {Zhang}}]{wang2015chiral}%
  \BibitemOpen
  \bibfield  {author} {\bibinfo {author} {\bibfnamefont {J.}~\bibnamefont {Wang}}, \bibinfo {author} {\bibfnamefont {Q.}~\bibnamefont {Zhou}}, \bibinfo {author} {\bibfnamefont {B.}~\bibnamefont {Lian}},\ and\ \bibinfo {author} {\bibfnamefont {S.-C.}\ \bibnamefont {Zhang}},\ }\bibfield  {title} {\bibinfo {title} {Chiral topological superconductor and half-integer conductance plateau from quantum anomalous hall plateau transition},\ }\href@noop {} {\bibfield  {journal} {\bibinfo  {journal} {Physical Review B}\ }\textbf {\bibinfo {volume} {92}},\ \bibinfo {pages} {064520} (\bibinfo {year} {2015})}\BibitemShut {NoStop}%
\bibitem [{\citenamefont {Bengtsson}(1999)}]{bengtsson1999dipole}%
  \BibitemOpen
  \bibfield  {author} {\bibinfo {author} {\bibfnamefont {L.}~\bibnamefont {Bengtsson}},\ }\bibfield  {title} {\bibinfo {title} {Dipole correction for surface supercell calculations},\ }\href@noop {} {\bibfield  {journal} {\bibinfo  {journal} {Physical Review B}\ }\textbf {\bibinfo {volume} {59}},\ \bibinfo {pages} {12301} (\bibinfo {year} {1999})}\BibitemShut {NoStop}%
\bibitem [{\citenamefont {Reja}\ \emph {et~al.}(2017)\citenamefont {Reja}, \citenamefont {Fertig}, \citenamefont {Brey},\ and\ \citenamefont {Zhang}}]{PhysRevB.96.201111}%
  \BibitemOpen
  \bibfield  {author} {\bibinfo {author} {\bibfnamefont {S.}~\bibnamefont {Reja}}, \bibinfo {author} {\bibfnamefont {H.~A.}\ \bibnamefont {Fertig}}, \bibinfo {author} {\bibfnamefont {L.}~\bibnamefont {Brey}},\ and\ \bibinfo {author} {\bibfnamefont {S.}~\bibnamefont {Zhang}},\ }\bibfield  {title} {\bibinfo {title} {Surface magnetism in topological crystalline insulators},\ }\href {https://doi.org/10.1103/PhysRevB.96.201111} {\bibfield  {journal} {\bibinfo  {journal} {Phys. Rev. B}\ }\textbf {\bibinfo {volume} {96}},\ \bibinfo {pages} {201111} (\bibinfo {year} {2017})}\BibitemShut {NoStop}%
\bibitem [{\citenamefont {Sattigeri}\ \emph {et~al.}(2023)\citenamefont {Sattigeri}, \citenamefont {Cuono},\ and\ \citenamefont {Autieri}}]{sattigeri2023altermagnetic}%
  \BibitemOpen
  \bibfield  {author} {\bibinfo {author} {\bibfnamefont {R.~M.}\ \bibnamefont {Sattigeri}}, \bibinfo {author} {\bibfnamefont {G.}~\bibnamefont {Cuono}},\ and\ \bibinfo {author} {\bibfnamefont {C.}~\bibnamefont {Autieri}},\ }\bibfield  {title} {\bibinfo {title} {Altermagnetic surface states: towards the observation and utilization of altermagnetism in thin films, interfaces and topological materials},\ }\href@noop {} {\bibfield  {journal} {\bibinfo  {journal} {Nanoscale}\ }\textbf {\bibinfo {volume} {15}},\ \bibinfo {pages} {16998} (\bibinfo {year} {2023})}\BibitemShut {NoStop}%
\bibitem [{\citenamefont {Haldane}(1988)}]{haldane1988model}%
  \BibitemOpen
  \bibfield  {author} {\bibinfo {author} {\bibfnamefont {F.~D.~M.}\ \bibnamefont {Haldane}},\ }\bibfield  {title} {\bibinfo {title} {Model for a quantum hall effect without landau levels: Condensed-matter realization of the "parity anomaly"},\ }\href {https://doi.org/10.1103/PhysRevLett.61.2015} {\bibfield  {journal} {\bibinfo  {journal} {Phys. Rev. Lett.}\ }\textbf {\bibinfo {volume} {61}},\ \bibinfo {pages} {2015} (\bibinfo {year} {1988})}\BibitemShut {NoStop}%
\bibitem [{\citenamefont {Liu}\ \emph {et~al.}(2016)\citenamefont {Liu}, \citenamefont {Zhang},\ and\ \citenamefont {Qi}}]{liu2016quantum}%
  \BibitemOpen
  \bibfield  {author} {\bibinfo {author} {\bibfnamefont {C.-X.}\ \bibnamefont {Liu}}, \bibinfo {author} {\bibfnamefont {S.-C.}\ \bibnamefont {Zhang}},\ and\ \bibinfo {author} {\bibfnamefont {X.-L.}\ \bibnamefont {Qi}},\ }\bibfield  {title} {\bibinfo {title} {The quantum anomalous hall effect: theory and experiment},\ }\href@noop {} {\bibfield  {journal} {\bibinfo  {journal} {Annual Review of Condensed Matter Physics}\ }\textbf {\bibinfo {volume} {7}},\ \bibinfo {pages} {301} (\bibinfo {year} {2016})}\BibitemShut {NoStop}%
\bibitem [{\citenamefont {Deng}\ \emph {et~al.}(2020)\citenamefont {Deng}, \citenamefont {Yu}, \citenamefont {Shi}, \citenamefont {Guo}, \citenamefont {Xu}, \citenamefont {Wang}, \citenamefont {Chen},\ and\ \citenamefont {Zhang}}]{deng2020quantum}%
  \BibitemOpen
  \bibfield  {author} {\bibinfo {author} {\bibfnamefont {Y.}~\bibnamefont {Deng}}, \bibinfo {author} {\bibfnamefont {Y.}~\bibnamefont {Yu}}, \bibinfo {author} {\bibfnamefont {M.~Z.}\ \bibnamefont {Shi}}, \bibinfo {author} {\bibfnamefont {Z.}~\bibnamefont {Guo}}, \bibinfo {author} {\bibfnamefont {Z.}~\bibnamefont {Xu}}, \bibinfo {author} {\bibfnamefont {J.}~\bibnamefont {Wang}}, \bibinfo {author} {\bibfnamefont {X.~H.}\ \bibnamefont {Chen}},\ and\ \bibinfo {author} {\bibfnamefont {Y.}~\bibnamefont {Zhang}},\ }\bibfield  {title} {\bibinfo {title} {Quantum anomalous hall effect in intrinsic magnetic topological insulator mnbi$_2$te$_4$},\ }\href@noop {} {\bibfield  {journal} {\bibinfo  {journal} {Science}\ }\textbf {\bibinfo {volume} {367}},\ \bibinfo {pages} {895} (\bibinfo {year} {2020})}\BibitemShut {NoStop}%
\bibitem [{\citenamefont {Serlin}\ \emph {et~al.}(2020)\citenamefont {Serlin}, \citenamefont {Tschirhart}, \citenamefont {Polshyn}, \citenamefont {Zhang}, \citenamefont {Zhu}, \citenamefont {Watanabe}, \citenamefont {Taniguchi}, \citenamefont {Balents},\ and\ \citenamefont {Young}}]{serlin2020intrinsic}%
  \BibitemOpen
  \bibfield  {author} {\bibinfo {author} {\bibfnamefont {M.}~\bibnamefont {Serlin}}, \bibinfo {author} {\bibfnamefont {C.}~\bibnamefont {Tschirhart}}, \bibinfo {author} {\bibfnamefont {H.}~\bibnamefont {Polshyn}}, \bibinfo {author} {\bibfnamefont {Y.}~\bibnamefont {Zhang}}, \bibinfo {author} {\bibfnamefont {J.}~\bibnamefont {Zhu}}, \bibinfo {author} {\bibfnamefont {K.}~\bibnamefont {Watanabe}}, \bibinfo {author} {\bibfnamefont {T.}~\bibnamefont {Taniguchi}}, \bibinfo {author} {\bibfnamefont {L.}~\bibnamefont {Balents}},\ and\ \bibinfo {author} {\bibfnamefont {A.}~\bibnamefont {Young}},\ }\bibfield  {title} {\bibinfo {title} {Intrinsic quantized anomalous hall effect in a moir{\'e} heterostructure},\ }\href@noop {} {\bibfield  {journal} {\bibinfo  {journal} {Science}\ }\textbf {\bibinfo {volume} {367}},\ \bibinfo {pages} {900} (\bibinfo {year} {2020})}\BibitemShut {NoStop}%
\bibitem [{\citenamefont {Li}\ \emph {et~al.}(2021)\citenamefont {Li}, \citenamefont {Jiang}, \citenamefont {Shen}, \citenamefont {Zhang}, \citenamefont {Li}, \citenamefont {Tao}, \citenamefont {Devakul}, \citenamefont {Watanabe}, \citenamefont {Taniguchi}, \citenamefont {Fu}, \citenamefont {Shan},\ and\ \citenamefont {Mak}}]{li2021quantum}%
  \BibitemOpen
  \bibfield  {author} {\bibinfo {author} {\bibfnamefont {T.}~\bibnamefont {Li}}, \bibinfo {author} {\bibfnamefont {S.}~\bibnamefont {Jiang}}, \bibinfo {author} {\bibfnamefont {B.}~\bibnamefont {Shen}}, \bibinfo {author} {\bibfnamefont {Y.}~\bibnamefont {Zhang}}, \bibinfo {author} {\bibfnamefont {L.}~\bibnamefont {Li}}, \bibinfo {author} {\bibfnamefont {Z.}~\bibnamefont {Tao}}, \bibinfo {author} {\bibfnamefont {T.}~\bibnamefont {Devakul}}, \bibinfo {author} {\bibfnamefont {K.}~\bibnamefont {Watanabe}}, \bibinfo {author} {\bibfnamefont {T.}~\bibnamefont {Taniguchi}}, \bibinfo {author} {\bibfnamefont {L.}~\bibnamefont {Fu}}, \bibinfo {author} {\bibfnamefont {J.}~\bibnamefont {Shan}},\ and\ \bibinfo {author} {\bibfnamefont {K.~F.}\ \bibnamefont {Mak}},\ }\bibfield  {title} {\bibinfo {title} {Quantum anomalous hall effect from intertwined moir{\'e} bands},\ }\href@noop {} {\bibfield  {journal} {\bibinfo  {journal} {Nature}\ }\textbf {\bibinfo {volume} {600}},\ \bibinfo {pages} {641} (\bibinfo {year}
  {2021})}\BibitemShut {NoStop}%
\end{thebibliography}%

\end{document}